\documentclass{iopjournal}

\usepackage{amsmath}
\usepackage{amssymb}
\usepackage{bbm}
\usepackage{bm}

\usepackage{comment}
\usepackage{ragged2e}

\usepackage{xcolor}

\definecolor{orange}{rgb}{1,0.5,0}
\definecolor{darkblue}{rgb}{0.,0.,0.4}
\definecolor{darkred}{rgb}{0.5,0.,0.}
\definecolor{BlueViolet}{RGB}{138,43,226}
\definecolor{SkyBlue}{RGB}{30,144,255}
\definecolor{DarkGreen}{RGB}{0,100,0}

\begin{document}

\articletype{Topical Review} 

\title{Quantum Anomalous Hall Effect in Rhombohedral Multilayer Graphene/hBN Moiré Superlattices}

\author{Jiannan Hua $^{1,2,*}$, 
Jing Ding$^{1,*}$, 
W. Zhu$^{1,2,\dagger}$, 
Shui-gang Xu$^{1,2,\#}$ 
}

\affil{$^1$ Department of Physics, School of Science, Westlake University, Hangzhou 310030, P. R. China}

\affil{$^2$ Institute of Natural Sciences, Westlake Institute for Advanced Study, Hangzhou 310024, P. R. China}

\affil{$^*$Authors equally contributes to this work.}

\email{zhuwei@westlake.edu.cn; xushuigang@westlake.edu.cn}



\begin{abstract}
The recent discovery of robust quantum anomalous Hall (QAH) effect in rhombohedral multilayer graphene (RMG) aligned with hexagonal boron nitride (hBN) has established a highly versatile platform for correlated topological matter. This review synthesizes the experimental and theoretical progress in understanding these interaction-driven topological phases. Experimentally, the landscape has rapidly expanded from initial Chern insulators in trilayer systems to fully quantized QAH states in thicker systems. 
Theoretically, it is believed that moiré potential and electron-electron interaction cooperate and produce the QAH effect in such systems.  
Theoretical calculations also bring interesting questions, such as the formation of an interaction-driven topological phase known as an anomalous Hall crystal (AHC). 
This review comprehensively covers the experimental hallmarks, the theoretical frameworks—including continuum models and many-body approaches—and the ensuing physical picture that reconciles the roles of interactions, displacement fields, and the moiré potentials. 
We conclude by outlining outstanding open questions and future directions, positioning RMG/hBN systems at the forefront of topological quantum matter.
\end{abstract}

\section{Introduction}

Recent experimental discovery of various Chern insulators in the rhombohedral multilayer graphene (RMG) aligned to hexagonal boron nitride (hBN) has provided a new platform for the study of correlated topological matter. 
Particularly, various topological states, including integer and fractional quantum anomalous Hall (QAH) effects, have been reported in these systems, stimulating the study of the interplay between strong interactions and intrinsic band topology. 
This review will be dedicated to reviewing the progress of the RMG/hBN moiré system in the past few years.

\subsection{Emergence of Moiré Systems}
The landscape of condensed matter physics has been revolutionized by the advent of two-dimensional (2D) materials, particularly graphene, with its unique Dirac band structure \cite{castro2009electronic}.
A significant paradigm shift occurred with the realization of moiré superlattices, formed by stacking and rotationally misaligning layers of graphene \cite{cao2018unconventional, cao2018correlated}, or graphene with other 2D materials such as hBN \cite{yankowitz2019tuning, kennes2021moire}. These moiré patterns introduce long-range periodic potentials that dramatically reconstruct the electronic band structure, leading to the emergence of flat electronic bands. Such flat bands enhance electron-electron interactions, fostering a rich array of exotic quantum phenomena, including correlated insulating states and unconventional superconductivity.

The experimental exploration of moiré systems gained significant momentum with twisted bilayer graphene (TBG). 
Seminal works by Cao \textit{et al.} \cite{cao2018unconventional,cao2018correlated} experimentally demonstrated unconventional superconductivity and correlated insulating states in TBG, validating earlier theoretical predictions of flat bands by Bistritzer and MacDonald \cite{bistritzer2011moire}. 
This breakthrough catalyzed extensive research into various twisted 2D homostructures and heterostructures \cite{yankowitz2019tuning}. 
The scope broadened considerably to include graphene aligned with hBN, where topological features and fractional quantum anomalous Hall (FQAH) states have been observed \cite{lu2024fractional,lu2025extended,xie2025tunable}, and to twisted transition metal dichalcogenides (TMDs) \cite{cai2023signatures, Zeng2023ThermodynamicFCItMoTe2, park2023observation, Xu2023PRXtMoTe2}. 
These pioneering studies collectively establish moiré systems as versatile platforms for realizing and manipulating a broad spectrum of quantum phases. 
Further advancements have quickly expanded along many different directions, such as FQAH with non-Abelian topological orders \cite{uchida2025non}. 
In this short review, we will mainly focus on the topological phases emergent in RMG/hBN superlattices. 

\subsection{Experiments in Moiré Superlattices}
The intricate interplay of strong correlations and topologies in moiré systems has led to an exceptionally rich landscape of novel quantum states, observed across both integer and fractional electronic fillings.

At integer fillings, a wide array of intriguing phenomena have been discovered. 
In few-layer rhombohedral graphene systems, including trilayer (3L), tetralayer (4L), pentalayer (5L), hexalayer (6L),  heptalayer (7L) and decalayer (10L) rhombohedral graphene aligned with hBN, researchers have observed a diverse correlated states. 
These include Mott insulator states \cite{chen2019evidence, chen2020tunable}, where strong electron-electron interactions localize charge carriers. Unconventional superconductivity in certain devices has also been a prominent finding \cite{choi2025superconductivity, kumar2025superconductivity, Matthew2025RGSC}. 
Orbital magnetism and various symmetry-broken states further enrich this phase space \cite{chen2022tunable, zhou2024layer}. 
Critically, Chern insulators and the QAH effect have been robustly observed at integer fillings in various moiré systems \cite{serlin2020intrinsic, Polshyn2020ElectricalSwitchingOrbitalChern, chen2020tunable, choi2025superconductivity, lu2025extended, lu2024fractional, xie2025tunable, xiang2025continuously, ding2025electric, Li2021QAHIntertwinedMoireBands}. 
These observations underscore the power of moiré engineering in inducing and controlling complex quantum states across different layer numbers. 

More interesting findings were at the fractional fillings, with the observation of fractional Chern insulators (FCIs) \cite{cai2023signatures, Zeng2023ThermodynamicFCItMoTe2, park2023observation, Xu2023PRXtMoTe2, lu2024fractional}. 
These states, analogous to the fractional quantum Hall (FQH) effect but occurring without external magnetic fields, represent a new frontier in topological quantum matter. Experimental reports confirm signatures of fractionally quantized anomalous Hall effects in various multilayer graphene systems \cite{lu2024fractional, lu2025extended, xie2025tunable}, highlighting the critical role of flat bands and strong correlations in stabilizing these exotic fractional topological phases. 
Capacitance measurements have been instrumental in these studies, providing insights into the incompressible states \cite{aronson2025displacement}. ARPES measurements also provide crucial insights into the electronic structure of these systems \cite{zhang2025moir}. 
However, the stability of these fractional states remains a significant area of ongoing research, particularly concerning the impact of multiband effects and various interaction schemes \cite{kwan2025moire, yu2025moire}.

\subsection{Theoretical Frameworks for Moiré Topological Systems}

Theoretical frameworks have been successful in interpreting the experimental discoveries in RMG/hBN moiré topological systems, offering predictive power and a deeper understanding of underlying mechanisms. 
Effective Hamiltonian models, such as continuum models, are widely employed to describe the long-wavelength moiré potentials arising from twist angles and lattice mismatches. 
These models successfully predict the dispersion relation with and without the moiré potentials \cite{koshino2009trigonal, wallbank2013generic, bokdam2014band, san2014spontaneous, moon2014electronic, jung2014ab, jung2015origin}.
Although Berry curvature can be finite and layer dependent within individual valleys, intrinsic RMG does not host isolated Chern bands. 
The introduction of a moiré potential breaks translational symmetry and reorganizes the band structure, giving rise to isolated Chern-band in RMG/hBN systems \cite{kumar2013flat, song2015topological, chittari2019gate, gonzalez2021topological, park2023topological, zhang2019nearly}.

After the intriguing experiment work \cite{lu2024fractional} showing the IQAH and FQAH in the RMG/hBN moiré system, 
to understand the correlated phenomena in flat-band moiré systems, many-body theoretical tools are crucial. Hartree-Fock (HF) and self-consistent Hartree-Fock (SCHF) calculations have been extensively used to predict spontaneous symmetry breaking, leading to the formation of isolated Chern bands, ferromagnetism, and other correlated phases \cite{zhang2019nearly, repellin2020ferromagnetism, dong2024anomalous, zhou2024fractional, dong2024theory, soejima2024anomalous, guo2024fractional, herzog2024moire, kwan2025moire, yu2025moire}.
The renormalization group (RG) method is employed to investigate this system \cite{guo2024fractional, liu2025layer}.
These methods provide critical insights into the energetic competition between different quantum states and the phase diagrams of moiré systems.

Furthermore, although most of the experimental observations can be largely reproduced using the theories that have been developed for the quantum Hall physics, the studies on the moiré systems also bring some unique insights. 
For example, the emergence of the anomalous Hall crystal (AHC) introduces a new theoretical paradigm, describing a topological electronic state spontaneously breaking time-reversal and continuous translational symmetries at zero magnetic field \cite{dong2024anomalous, zhou2024fractional, dong2024theory, guo2024fractional, herzog2024moire}.
Subsequently, researchers have been attracted to interest in AHC, including the conditions for its formation, its stability, and its competition with other correlated topological phases \cite{soejima2024anomalous, kwan2025moire, yu2025moire, dong2024stability, zeng2024sublattice, huang2024self, soejima2025lambda}. 
This phase provides a novel mechanism for generating Chern bands, enriching the theoretical landscape.

\subsection{Structure of this Review}
This review contains five sections. In Sec. \ref{sec:experiment}, we will review the experimental discoveries of QAH in RMG/hBN systems by highlighting several hallmarks observed in the experiments. In Sec. \ref{sec:calculation}, we review the theoretical basis to describe this RMG/hBN system. In Sec. \ref{sec:theory}, we discuss the recent theoretical understanding on the QAH observed in this system. 
In Sec. \ref{sec:discussion}, we close the discussion and present some open directions.

\section{Experimental discoveries of QAH in RMG/hBN moir\'e superlattices}
\label{sec:experiment}

The experimental realization of the QAH effect in RMG emerged progressively through studies spanning different layer numbers and electrostatic tuning regimes. 
Rather than appearing abruptly, QAH physics in these systems initially developed from interaction-driven Chern insulating behavior \cite{chen2020tunable,chen2022tunable} and evolved into robust zero-field quantized Hall states \cite{lu2024fractional, lu2025extended, xie2025tunable, ding2025electric} after the roles of moir\'e superlattices, displacement fields, and spontaneous symmetry breaking became clear. 
In this section, we summarize the key experimental advances that established RMG/hBN systems as a platform for QAH and related topological phases. 

\subsection{Chern insulators in R3G/hBN moir\'e superlattices}
The low-energy electronic structure of multilayer graphene depends strongly on layer number and stacking order. 
Rhombohedral trilayer graphene (R3G) hosts a cubic band dispersion, which enhances the density of states and electron--electron interactions compared to monolayer and bilayer graphene. 
When aligned with hexagonal boron nitride, the additional moiré band folding further amplifies correlation effects \cite{koshino2009trigonal,koshino2010interlayer,zhang2010band,min2008chiral}.

Early transport experiments on R3G/hBN moir\'e superlattices revealed interaction-driven insulating states at integer moir\'e fillings, accompanied by sizable anomalous Hall effect \cite{chen2020tunable, chen2022tunable, Guorui2022review}. 
Although the Hall resistance was not fully quantized at zero magnetic field, its evolution under small perpendicular fields followed the Streda relation \cite{streda1982theory}, yielding well-defined integer Chern numbers \cite{chen2020tunable}. 
These observations demonstrated the existence of Chern insulators stabilized by interactions induced valley polarization, indicating spontaneous breaking of time-reversal symmetry \cite{chen2020tunable}. 
However, the need for a finite magnetic field to fully resolve quantized plateaus suggests that these states represent precursors rather than fully developed QAH phases. 
Nevertheless, the R3G/hBN moir\'e superlattice results provided a crucial proof of principle that RMG/hBN moir\'e systems can host interaction-driven topological bands without Landau level quantization \cite{chen2020tunable,chen2022tunable}.

\subsection{Integer and fractional QAH in RMG/hBN moir\'e superlattices}

\begin{figure}
 \centering
\includegraphics[width=0.75\textwidth]{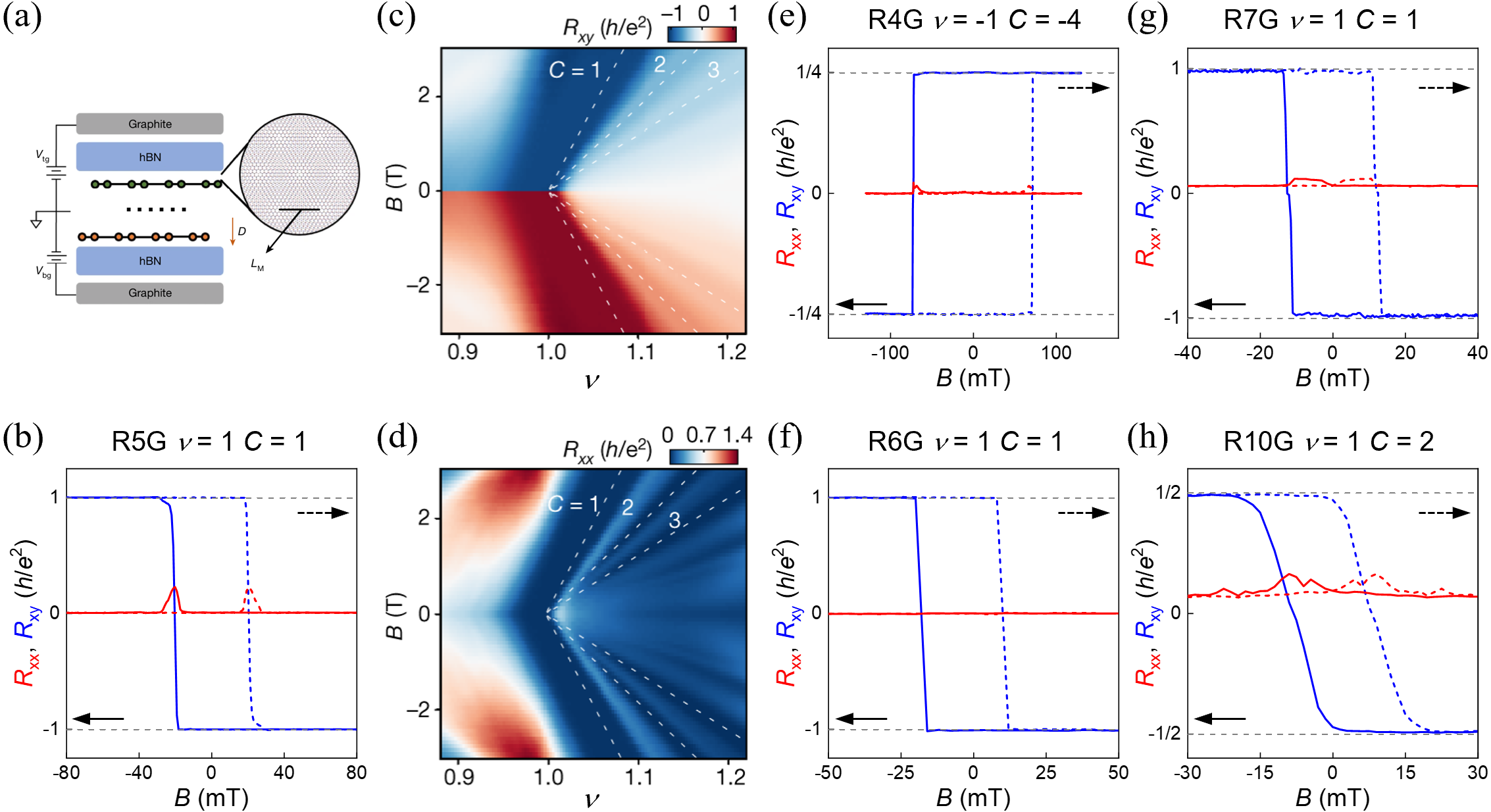}
 \caption{
 IQAH in RMG/hBN moir\'e superlattices. 
  (a) Device schematic and measurement configuration of a standard dual-gate structure (reproduced from Ref. \cite{lu2025extended}). 
   (b) Magnetic hysteresis of $R_{xy}$ (blue) and $R_{xx}$ (red) at $\nu=1$ and $|D|=0.97~\mathrm{V\,nm^{-1}}$ in R5G/hBN moiré  superlattices. 
   At $200 ~\mathrm{mK}$,  $R_{xy}$ is quantized at $h/e^{2}$ consistent with a Chern number $C=1$, whereas $R_{xx}$ is strongly suppressed at zero field. 
Solid (dashed) traces correspond to sweeping $B$ backward (forward) (replotted from data in Ref. \cite{lu2024fractional}). 
   (c),(d) Landau fan diagrams of $R_{xy}$ (c) and $R_{xx}$ (d) at $|D|=0.97~\mathrm{V\,nm^{-1}}$ in R5G/hBN moiré  superlattices. The IQAH state forms a wide plateau whose Landau-fan slope matches the $C=1$ dashed line expected from the Streda formula (reproduced from Ref. \cite{lu2024fractional}). 
   (e)-(h) Representative IQAH responses at integer moiré filling in different layer numbers: 
   (e) R4G at $\nu=-1$ and $|D|=0.21~\mathrm{V\,nm^{-1}}$ with $C=-4$ (replotted from data in Ref. \cite{choi2025superconductivity}), 
   (f) R6G at $\nu=1$ and $|D|=0.51~\mathrm{V\,nm^{-1}}$ with $C=1$ (replotted from data in Ref. \cite{xie2025tunable}), 
   (g) R7G  at $\nu=1$ and $|D|=0.64~\mathrm{V\,nm^{-1}}$ with $C=1$ (from Ref. \cite{xiang2025continuously}),  
   (h) R10G at $\nu=1$ and $|D|=0.81~\mathrm{V\,nm^{-1}}$ with $C=2$ (from Ref. \cite{ding2025electric}). Moiré filling factors ($\nu$) and Chern numbers ($C$) are labeled above each panel. Solid (dashed) traces correspond to sweeping $B$ backward (forward). 
}
\label{fig:exp1_IQAH_RMG}
\end{figure}

Increasing the number of rhombohedral layers further enhances correlation effects through the \(k^{N}\) low-energy dispersion where $k$ is the wave number \cite{shi2020electronic,koshino2009trigonal,koshino2010interlayer} . 
A major breakthrough was achieved in rhombohedral pentalayer graphene (R5G) aligned with hBN, where fully quantized anomalous Hall states were observed at zero magnetic field \cite{lu2024fractional}. 
At moir\'e filling \(\nu = 1\), transport measurements showed a quantized Hall resistance \(\,R_{xy} = \pm h/e^{2}\,\) together with vanishing longitudinal resistance \(\,R_{xx}\,\) \cite{lu2024fractional}, as is shown in Fig.~\ref{fig:exp1_IQAH_RMG}(b). A characteristic feature of the QAH states in R5G/hBN moir\'e superlattice is pronounced magnetic hysteresis in \(R_{xy}\), providing direct evidence for spontaneous magnetization and broken time-reversal symmetry \cite{lu2024fractional}. 
The Streda formula analysis further supports an integer-Chern insulating state, consistent with the standard experimental criteria for the integer QAH effect, as shown in Fig.~\ref{fig:exp1_IQAH_RMG}(c)(d).
These results established rhombohedral graphene moir\'e superlattices as a clean and reproducible platform for interaction-driven QAH physics.

\begin{figure}
 \centering
\includegraphics[width=0.75\textwidth]{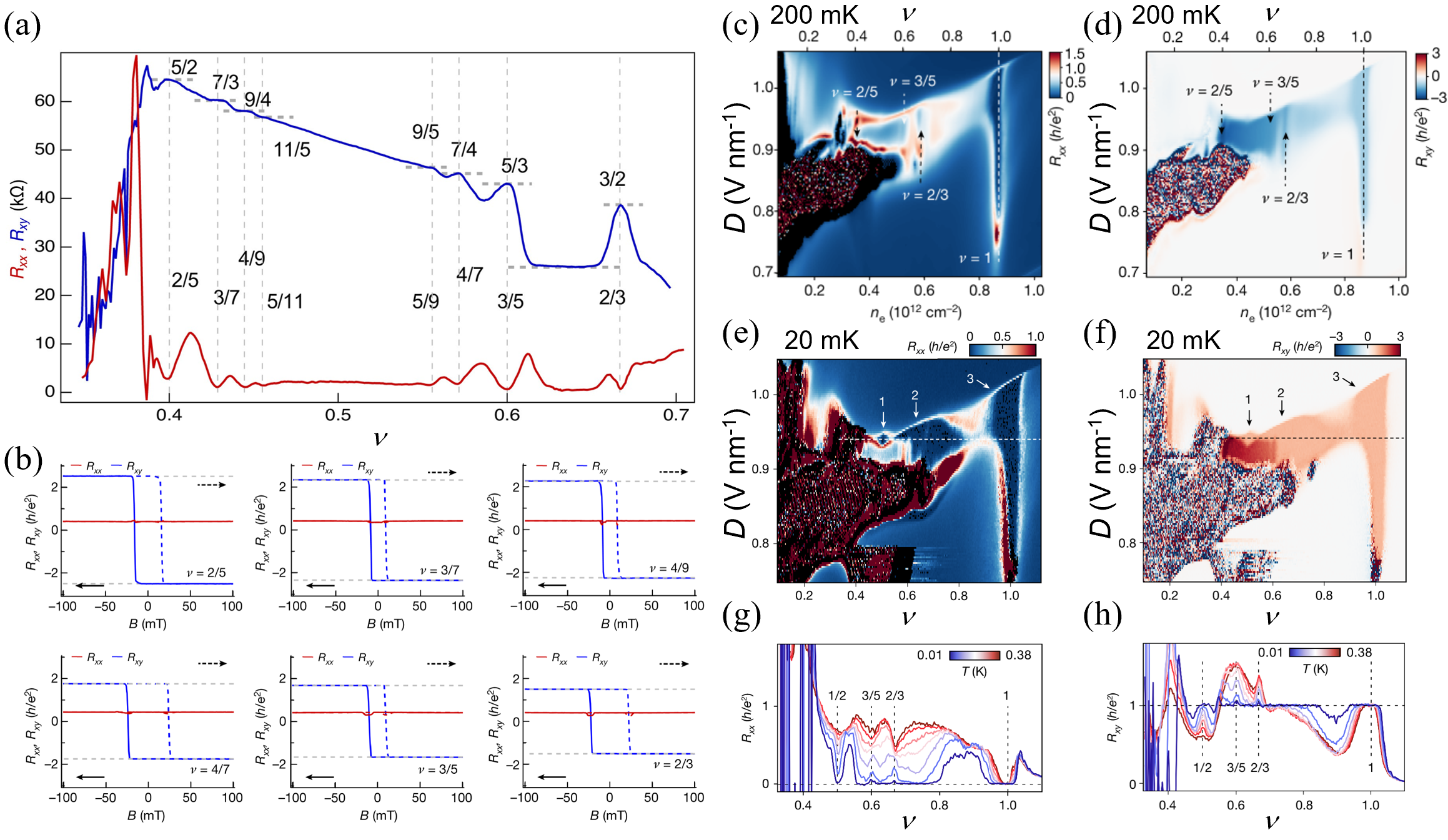}
  \caption{
 Fractional and extended QAH in R5G/hBN moir\'e superlattices.
 (a) $R_{xy}$ (blue) and $R_{xx}$ (red) as functions of filling factor. $R_{xy}$ plateaus and $R_{xx}$ dips occur at  fractional fillings $\nu=2/5, 3/7, 4/9, 5/11, 5/9, 4/7, 3/5$ and $2/3$, at $20~\mathrm{mK}$. 
(b) Representative magnetic hysteresis scans of $R_{xy}$ and $R_{xx}$ at selected fractional fillings $\nu=2/5, 3/7, 4/9, 4/7, 3/5$ and $2/3$, showing quantized \(R_{xy} = h/(\nu e^{2})\) with vanishing $R_{xx}$. 
(c),(d) Phase diagrams of $R_{xx}$ (c) and $R_{xy}$ (d) at $B=\pm0.1~\mathrm{T}$ as functions of filling factor $\nu$ and displacement field $D$, taken at electron temperature of $200~\mathrm{mK}$ , where $R_{xy}$ plateaus and $R_{xx}$ dips occur in a stripe-like region and fractional features appear at commensurate $\nu=1, 2/3, 3/5$ and $2/5$ (indicated by the dashed lines and arrows). 
(e),(f) Extended phase diagrams of $R_{xx}$ (e) and $R_{xy}$ (f) at $B=\pm0.1~\mathrm{T}$ as functions of $\nu$ and $D$, taken at $20~\mathrm{mK}$. Three regions (labelled by numbers and arrows) show quantized $R_{xy}$ at $h/e^2$ and vanishing  $R_{xx}$, over a wide window (extended QAH regime). The dashed lines correspond to $|D|=0.95~\mathrm{V\,nm^{-1}}$. 
(g),(h) Temperature-dependent $R_{xx}$ (g) and $R_{xy}$ (h) taken along the dashed line in (e) and (f), showing how the extended quantized region evolves and competes with nearby fractional states. 
All data were extracted by a symmetrizing/anti-symmetrizing process under magnetic-field reversal, $R_{xx}^{\mathrm{sym}}(\pm B)=\frac{R_{xx}(+B)+R_{xx}(-B)}{2}$, $R_{xy}^{\mathrm{asym}}(\pm B)=\frac{R_{xy}(+B)-R_{xy}(-B)}{2}$. 
Data in panel (a) are from pentalayer device D2; all other panels [(b)-(h)] are from device D1.
(a),(e)-(h) are reproduced from Refs.~\cite{lu2025extended}. (b)-(d) are reproduced from Refs.~\cite{lu2024fractional} with permission. \copyright 2024, 2025 Springer Nature.}
\label{fig:exp2_FQAH&EQAH_RMG}
\end{figure}

Beyond integer filling, R5G/hBN moir\'e superlattices exhibit a qualitatively new regime: the FQAH effect \cite{lu2024fractional,lu2025extended}. 
At zero magnetic field, a series of fractional Hall plateaus was observed at fractional moir\'e fillings, including \(\nu = 2/5, 3/7, 4/9, 5/11, 5/9, 4/7, 3/5\) and \(2/3\), with Hall resistance quantized to \(R_{xy} = h/(\nu e^{2})\) and a concurrent suppression of \(R_{xx}\) \cite{lu2024fractional,lu2025extended}, as shown in Fig.~\ref{fig:exp2_FQAH&EQAH_RMG}(a)(b). 
These observations constitute the first realization of FQAH in graphene.

The fractional states arise from topological flat bands and are stabilized by strong Coulomb interactions, rather than by Landau level physics \cite{lu2024fractional, lu2025extended, aronson2025displacement}. 
In addition, an extended QAH (EQAH) regime has been reported at electronic temperature of $20~\mathrm{mK}$, as shown in Fig.~\ref{fig:exp2_FQAH&EQAH_RMG}(e)(f), where the Hall quantization persists over a wide carrier-density window rather than being pinned to a single discrete filling \cite{lu2025extended}.
(See Sec. \ref{sec:discussion}  for a detailed discussion of the mechanism underlying EQAH.) 
In R5G/hBN moiré devices, FQAH states are frequently observed adjacent to IQAH and EQAH regimes, as shown in Fig.~\ref{fig:exp2_FQAH&EQAH_RMG}(c)(d). This proximity, together with the disorder-tuned replacement between FQAH and EQAH states, indicates strong competition between these two phases. \cite{lu2024fractional,lu2025extended}.

Motivated by the results observed in pentalayer graphene, IQAH have also been reported in other RMG/hBN moir\'e devices with different layer numbers $N$ at integer moir\'e fillings, including \(\nu = -1\) ($N=4$), \(1\) ($N=4 \sim 8,10$) and \(2\) ($N=7,8$, at large $D$) \cite{choi2025superconductivity,xie2025tunable,xiang2025continuously,Jianming2026R7GPRL,kumar2025superconductivity,xiaobo2025R8GarXiv,ding2025electric}.
FQAH effect has also been observed in rhombohedral tetralayer graphene (R4G) and hexalayer graphene (R6G) aligned with hBN, demonstrating that FQAH physics is not limited to pentalayer alone \cite{lu2025extended,xie2025tunable}. 
The observation of FQAH signatures across R4G, R5G, and R6G establishes fractional topology as a generic emergent phenomenon in RMG moir\'e systems \cite{lu2024fractional,lu2025extended,xie2025tunable}.

\subsection{High-Chern-number QAH and chirality switching in RMG/hBN moir\'e superlattices}

The Chern number of QAH insulators characterizes both the number of chiral edge channels and their propagation direction. 
Thicker rhombohedral multilayers enable electrical control over both the magnitude and sign of the Chern number \cite{ding2025electric,xiang2025continuously,wang2025electrical,xie2025tunable,Guorui2025R6GPRL,Xiaobo2025R6Ginplane,Jianming2026R7GPRL}.

In rhombohedral heptalayer graphene (R7G) and decalayer graphene (R10G) aligned with hBN, robust Chern insulating states with \(C = 2\) have been observed \cite{ding2025electric,wang2025electrical}. 
In particular, zero-field quantization of the \(C = 2\) QAH state was demonstrated in R10G moir\'e superlattices proximitized by WSe\(_2\) \cite{ding2025electric}, as shown in Fig.~\ref{fig:exp1_IQAH_RMG}(h). 
Transport measurements reveal strong competition among Chern insulating phases with \(C = -1, 1\) and \(2\), controlled by the displacement field \cite{ding2025electric}. 
At fixed filling, displacement field can modulate Berry distribution of the topological band and therefore induce topological phase transitions that change both the magnitude and sign of the Chern number, resulting in electrically controlled chirality switching near zero magnetic field \cite{ding2025electric,xiang2025continuously,wang2025electrical}.

In addition, a higher-Chern-number topological state appears in the second correlated band over $1.5<\nu<2$, where a Streda analysis of the Landau-fan slope gives $C=3$ \cite{xiang2025continuously}.
By contrast, the $\nu=2$ state at similar $D$ is identified to be topologically trivial state ($C=0$), showing large $R_{xx}$,
non-dispersive $R_{xy}$, and the absence of anomalous Hall signals \cite{xiang2025continuously}.
Importantly, the $C=3$ phase is stabilized only within a finite pocket in the $(\nu,D)$ phase diagram:
its intercept filling shifts continuously with displacement field $D$ \cite{xiang2025continuously}.
These observations support the central role of interactions and incommensurate electronic
crystallization in selecting the Chern states, with $D$ acting as a primary control parameter
that modifies the band structure and Berry curvature distribution \cite{xiang2025continuously}.

\subsection{Influence of the moir\'e potential on QAH}

Here we define two gate-selectable regimes, termed ``moir\'e-proximal'' and ``moir\'e-distant''.
They refer to carriers --- specifically the electrons of filled conduction bands --- distribution that are predominantly polarized toward or away from the graphene/hBN interface, respectively \cite{aronson2025displacement}. 
It's noted that in some studies, the term “moiré-proximal” is also referred to as “moiré-proximate” \cite{dong2024anomalous}.
In several RMG devices, the most robust QAH and related Chern insulators appear in the moir\'e-distant regime \cite{lu2024fractional, lu2025extended, choi2025superconductivity, ding2025electric, xiang2025continuously, xie2025tunable}. 
Although topological states also have been reported in the moiré proximal side in R5G/hBN moiré superlattices, these states need a magnetic field to be stabilized \cite{lu2024fractional, lu2025extended, aronson2025displacement, choi2025superconductivity, ding2025electric, xie2025tunable, Eli2025orientation, Xiaoxue2025orientation}. 
This challenges the conventional expectation that a strong moir\'e potential is always required to stabilize isolated topological flat bands \cite{bistritzer2011moire,moon2014electronic,jung2014ab,jung2015origin,jung2017moire}, and instead suggests that interaction-driven valley polarization and intrinsic band topology can independently generate Chern bands, with the moir\'e potential acting primarily as a tunable symmetry-breaking perturbation \cite{lu2024fractional,lu2025extended,ding2025electric,JCCMP2024Aug}.

With this moir\'e-proximal/moir\'e-distant contrast in mind, it is useful to highlight two experimental knobs that tune the moir\'e landscape: the twist angle and the alignment orientation. The twist angle modulates the moir\'e wavelength and therefore sets the overall scale of moir\'e band folding and reconstruction, which is most directly reflected in the correlated features on the moir\'e-proximal side \cite{Eli2025orientation}. 
In the twist-angle range  currently explored, smaller twist angles correlate with a broader displacement-field stability window for the moir\'e-distant ($\nu=1$, $C=1$) QAH phase; as the angle $\theta$ increases, the critical displacement field $D_c$ shifts to higher values and the stability window narrows, with the $C=1$ QAH state collapsing by $\theta \approx 1.10^\circ$ at zero magnetic field. Moreover, FQAH states at $\nu=2/3$ exhibit even stronger angular sensitivity and are only resolved in sufficiently small-angle devices \cite{Xiaobo2025orientation}.

The alignment orientation provides an additional handle. 
Available data indicate that changing the orientation can noticeably reshape correlated phenomena in the moir\'e-proximal regime, while having much weaker impact on the emergence of integer and fractional Chern insulators in the moir\'e-distant regime \cite{Eli2025orientation, Xiaobo2025orientation, Xiaoxue2025orientation}. 
In practice, this is consistent with the view that moir\'e-distant Chern phases are mainly selected by interactions within the intrinsic rhombohedral bands, whereas the moir\'e potential more strongly affects the competition between nearby states and the size of the stability windows \cite{lu2024fractional, lu2025extended, Eli2025orientation, Xiaobo2025orientation, Xiaoxue2025orientation, JCCMP2024Aug}.

\subsection{Summary of experimental trends}

In summary, experiments on RMG have established a broad landscape of QAH
phenomena across multiple layer numbers and electrostatic tuning regimes.
Chern insulators were first identified in R3G, while robust zero-field integer and fractional
QAH states were subsequently realized in R5G and also observed in neighboring multilayers such as R4G and R6G.
In thicker systems (e.g., R7G and R10G), high-Chern-number QAH states and gate-tunable chirality have been
reported.
Rather than implying a strictly monotonic or universal layer-number dependence, the available results highlight how interaction strength, layer polarization, and band topology jointly shape the accessible phases, with layer number serving as one important tuning axis among several.
These observations establish RMG/hBN moir\'e systems as a versatile platform for exploring interaction-driven topological matter, and motivate a closer examination of open questions and broader implications in the Sec. \ref{sec:discussion}.

\section{Theoretical description of RMG/hBN moiré system}
\label{sec:calculation}

This section delves into the specific characteristics of RMG/hBN moiré system, its theoretical descriptions, and the underlying mechanisms by which its unique properties lead to rich topological physics.

\subsection{RMG's crystal structure}

\subsubsection{Monolayer graphene}

We begin by briefly reviewing the lattice structure of monolayer graphene, which serves as the fundamental building block of all multilayer graphene systems. 
A graphene monolayer forms a two-dimensional honeycomb lattice with lattice constant $a_G \approx 2.46\,${\AA} consisting of two inequivalent sublattices labeled A and B, as shown in figure \ref{fig:lattice_and_bands}(a).

The real-space primitive lattice vectors are given by
\begin{equation}\label{}
\begin{aligned}
    \bm{R}_1 &= a_G \left(1, 0 \right),\text{\quad} \bm{R}_2=a_G \left(-\frac{1}{2}, \frac{\sqrt{3}}{2} \right) \\
\end{aligned}
\end{equation}
with corresponding reciprocal lattice vectors
\begin{equation}\label{}
\begin{aligned}
    \bm{g}^{\text{G}}_1 = \frac{2\pi}{a_G} \left(1, \frac{1}{\sqrt{3}} \right), \text{\quad} \bm{g}^{\text{G}}_2 = \frac{2\pi}{a_G} \left(0, \frac{2}{\sqrt{3}} \right)
\end{aligned}
\end{equation}
The intra-layer nearest-neighbor bonds connecting the A and B sublattices are described by the vectors $\bm{\delta}_1 = \frac{a_G}{\sqrt{3}}\left(0, 1 \right)$, $\bm{\delta}_2 = \frac{a_G}{\sqrt{3}}\left(-\frac{\sqrt{3}}{2}, -\frac{1}{2} \right)$ and $\bm{\delta}_3 = \frac{a_G}{\sqrt{3}}\left(\frac{\sqrt{3}}{2}, -\frac{1}{2} \right)$.

\subsubsection{RMG}

\begin{figure}
 \centering
        \includegraphics[width=0.95\textwidth]{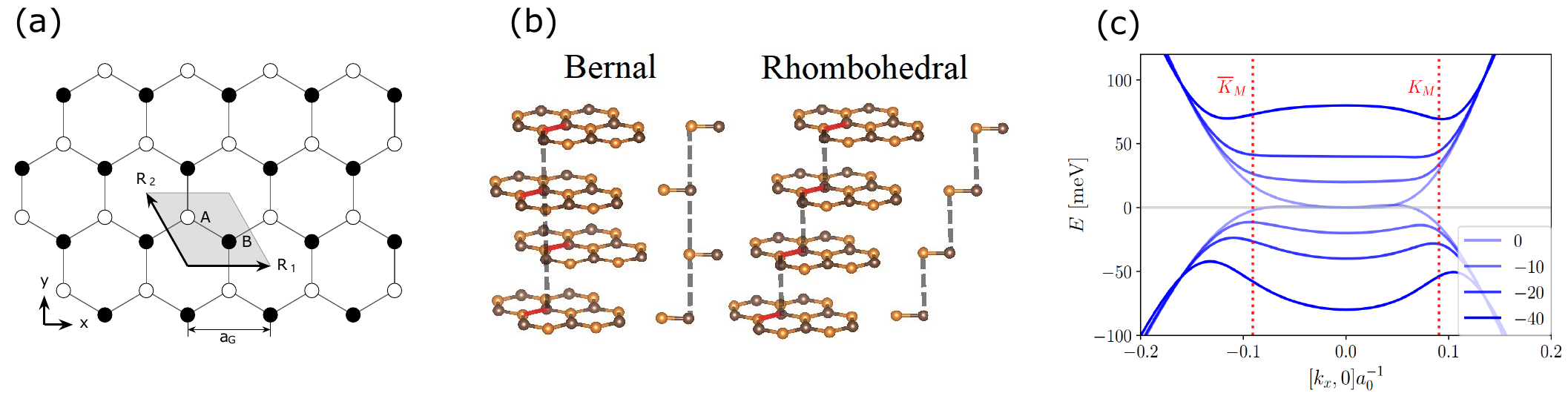}
 \caption{\textbf{Lattice structure and displacement-field-tunable band structures of RMG.} 
    (a). Top view of the honeycomb lattice. For the carbon-carbon bonds along the $y$-axis, the atoms at the lower and upper positions are defined as A (white circles) and B (black circles) sublattices, respectively. $\boldsymbol{R}_1$ and $\boldsymbol{R}_2$ denote the primitive lattice vectors, and $a_G$ is the carbon-carbon bond length. 
    (b). Schematic side view of stacking configurations comparing Bernal (AB) and rhombohedral (ABC) stacking. The dashed lines indicate the vertical alignment and shifting of sublattices across layers. 
    (c). (Cited from \cite{dong2024theory}) Continuum pentalayer rhombohedral graphene dispersion (without hBN alignment) along the $k_x$ axis for increasing interlayer potential difference $\Delta$ (meV) highlighted by darker shades of blue. The vertical dashed lines indicate the location of the moiré $K$ points  when h-BN is aligned with pentalayer graphene.}
    \label{fig:lattice_and_bands}
\label{fig:stack}
\end{figure}

Multilayer graphene structures are generated by stacking graphene monolayers with specific relative lateral translations. 
The two most common stacking geometries are Bernal (ABA) stacking and rhombohedral (ABC) stacking \cite{shi2020electronic, zhou2024layer}, as illustrated in figure \ref{fig:lattice_and_bands}(b).
In Bernal stacking, adjacent layers alternate between two registry configurations, forming an ABAB... sequence. 
In contrast, rhombohedral graphene exhibits an ABC stacking order, in which each successive layer is shifted relative to the previous one in the same direction. 
This stacking pattern can be visualized as a staircase-like structure, where the fourth layer is positioned directly above the first layer, completing the three-layer ABC stacking period. 
As a result, the crystallographic unit cell of ABC-stacked graphene differs fundamentally from that of Bernal-stacked graphene, and inversion symmetry present in ABA structures is broken.

In this work, we focus on RMG consisting of $N$-layer graphene stacked in the ABC configuration. Each layer contributes two carbon atoms per unit cell, labeled as $A_l$ and $B_l$, where 
$l\in [1, N]$ denotes the layer index. 
The positions of these atoms within the unit cell are
\begin{equation}\label{}
\begin{aligned}
    \bm{r}_{A_l} = \left(0, \frac{l-1}{\sqrt{3}}a_G, c\right), \text{\quad} \bm{r}_{B_l} = \left(0, \frac{l}{\sqrt{3}}a_G, c\right)
\end{aligned}
\end{equation}
where $c$ is the interlayer spacing, whose precise value is unimportant for the low-energy theory.

Each carbon atom provides one unhybridized $p_z$ orbital.
The corresponding second-quantized fermionic creation operator is defined as
$\hat{c}^\dagger_{\bm{R},\alpha}=\hat{c}^\dagger(\bm{R} + \bm{r}_{\alpha})$, where $\bm{R}=n_1 \bm{R}_1 + n_2 \bm{R}_2$ is the lattice vector with integers $n_1$, $n_2$ and $\alpha \in \{ A_l, B_l \}$ corresponds to the mixture of sublattice and layer indices.
In the subsequent discussion, any mention of ``sublattice" of RMG implicitly includes the layer index, unless stated otherwise.
The Fourier-transformed operator is $\hat{c}^\dagger_{\bm{k},\alpha}= \frac{1}{\sqrt{\mathcal{N}}} \sum_{\bm{R}} e^{-i\bm{k}\cdot (\bm{R}+\bm{r}_{\alpha})}\hat{c}^\dagger_{\bm{R}, \alpha}$, with normalization coefficient $\frac{1}{\sqrt{\mathcal{N}}}$ such that it obeys canonical anticommutation relations $\{ \hat{c}_{\bm{k}, \alpha}, \hat{c}_{\bm{k}, \alpha^\prime} \} = \delta_{\alpha, \alpha^{\prime}}$.

\subsection{Tight-Binding Models and Single-Particle Band Structure}

RMG exhibits a highly nontrivial low-energy electronic structure that depends sensitively on both layer number and external perturbations, such as a perpendicular displacement field \cite{zhou2021superconductivity, zhou2021half, liu2023interaction, han2024correlated, shi2020electronic}.
To quantitatively capture these features, it is necessary to adopt a microscopic description that retains lattice-scale hopping processes while remaining amenable to systematic low-energy analysis.

The effective tight-binding Hamiltonian for $N$-layer rhombohedral graphene can be expressed as $H_{N} = \sum_{\bm{k}} H_{N}(\bm{k})$, where $H_{N}(\bm{k})$ is a matrix in momentum space that encodes the coupling between sublattice and layer degrees of freedom.
Depending on the level of approximation and the physical phenomena of interest, $\hat{H}_N(\bm{k})$ can take different explicit forms, which will be introduced below.

\subsubsection{Full-site Hamiltonian}

A straightforward approach to studying the single-particle band structure of RMG is to construct its effective Hamiltonian based on the Slonczewski-Weiss-McClure (SWMC) tight-binding lattice model \cite{slonczewski1958band, mcclure1957band, zhou2024fractional, dong2024theory, dong2024anomalous, park2023topological}. 
In the basis $\hat{c}_{\bm{k}} = (\hat{c}_{\bm{k}, A_1}, \hat{c}_{\bm{k}, B_1}, \dots, \hat{c}_{\bm{k}, A_N},  \hat{c}_{\bm{k}, B_N})^\top$, 
the Hamiltonian of $N$-layer rhombohedral graphene is given by
\begin{equation}
    \hat{H}_N(\bm{k}) =
    \hat{c}_{\bm{k}}^\dagger
    \begin{pmatrix}
     D_1       &  V        & W       & 0        & 0   \\
     V^\dagger &  D_2      &  V       &  \ddots & 0    \\
     W^\dagger & V^\dagger &  \ddots  &   \ddots & W   \\
      0        &  \ddots   &  \ddots  &  D_{N-1}  & V  \\
      0        &   0       &  W^\dagger & V^\dagger & D_{N}  \\
    \end{pmatrix}
\hat{c}_{\bm{k}}
\end{equation}
where $2\times2$ blocks are defined as
\begin{equation}\label{eq:H_N_block}
    D_l = 
    \begin{pmatrix}
     u_{A_l}     &  -t_0 f(\bm{k})     \\
     -t_0  f^{*}(\bm{k})   &  u_{B_l}      \\
    \end{pmatrix} + \delta_l \mathbbm{1}  , \text{\quad}
    V = 
    \begin{pmatrix}
     t_4 f(\bm{k})   & t_3  f^{*}(\bm{k})    \\
     t_1                 & t_4 f(\bm{k})     \\
    \end{pmatrix}, \text{\quad}
    W = 
    \begin{pmatrix}
     0  & \frac{t_2}{2}    \\
     0  & 0   \\
    \end{pmatrix}.
\end{equation}
Here, $t_0=2.6\,\mathrm{eV}$ and $t_1=0.3561\,\mathrm{eV}$ describe nearest-neighbor intra-layer and inter-layer hopping, respectively.
$t_3=0.293\,\mathrm{eV}$ corresponds to hopping between sites $A_l \leftrightarrow B_{l+1}$, while $t_4=0.144\,\mathrm{eV}$ couples $A_l \leftrightarrow A_{l+1}$ and also $B_l \leftrightarrow B_{l+1}$.
$t_2=-0.0166\,\mathrm{eV}$ denotes direct hopping between $A_l \leftrightarrow B_{l+2}$ sites, although some works incorporate $t_2$ directly as the matrix element of $W$ \cite{park2023topological, kwan2025moire}.

The on-site potential difference
\begin{equation}
    u_{\alpha} = 
\begin{cases}
    0  \, \mathrm{meV}, & \alpha=A_1, B_N \\
    12.2 \, \mathrm{meV}, & \alpha=A_N, B_1  \\
    -16.4 \, \mathrm{meV}, & \text{otherwise}. \\
\end{cases}
\end{equation}
originates from variations in the atomic microenvironments.
The layer-dependent on-site potential energy $\delta_l=(\frac{1+N}{2}-l)\Delta$ models the potential energy difference between adjacent graphene layers induced by an external perpendicular electric (displacement) field.
The structure factor $f(\bm{k}) = \sum_{j=1}^3 e^{i \bm{k}\cdot \bm{\delta}j}$ can be expanded near the corners of the Brillouin zone $\bm{K}{\tau=\pm} = \left(\tau \frac{4\pi}{3a_G}, 0\right)$ as
\begin{equation}
f(\bm{K}_{\tau} + \bm{k}) \approx f(\bm{K}_{\tau}) + \nabla_{\bm{k}} f(\bm{K}_{\tau}) \cdot \bm{k} = -\frac{\sqrt{3}}{2}a_G(\tau k_x - i k_y)
\end{equation}
Thus, for momentum $\bm{p} = (p_x, p_y)= \hbar \bm{k}$ measured from $\bm{K}_{\tau}$, one may replace $t_i f(\bm{k})$ by $-v_i \pi^{*}$, where the Fermi velocity parameters $v_i=\frac{\sqrt{3}a_G}{2\hbar} t_i$ and $\pi=\tau p_x + i p_y = \hbar(\tau k_x + i k_y)$ \cite{park2023topological}.
This continuum expansion forms the basis for deriving low-energy effective models of RMG.
Some may omit $\hbar$ in both definitions of $v_i$ and $\pi$, and absorb $a_G$ into $\bm{k}$ for convenience. Hopping parameters $t_i$ can vary across different studies. 

To be more accurate, some studies \cite{herzog2024moire} add a inversion symmetric chemical potential $[H_{\text{ISP}}]_{ll'} = V_{\text{ISP}}\delta_{ll'}|l - \frac{N+1}{2}|$ where $V_{\text{ISP}} = 16.65 \, \mathrm{meV}$ indicates the potential of direction internal symmetrical polarization due to the different chemical environment of outer and inner atoms in the graphene.
But this term is not included in this review for simplicity.


\subsubsection{Two-site Hamiltonian}
Beyond the full-site Hamiltonian, an effective two-site Hamiltonian, obtained by projecting the full lattice Hamiltonian onto the non-dimerized sites $A_1$ and $B_N$ on the bottom and top layers, has been widely adopted in previous studies to describe the low-energy electronic structure of RMG \cite{chittari2019gate, zhang2019nearly, gonzalez2021topological, patri2023strong, chatterjee2022inter}, where the low-energy bands are predominantly contributed by these two sublattices.
Near the Fermi level (the charge neutrality point), the electronic states of the two lowest-energy bands are predominantly localized on the $A_1$ and $B_N$ sites.
In particular, at $\bm{k}=0$ these two low-energy eigenstates are exactly localized on the $A_1$ and $B_N$ sites.
This localization originates from the strong interlayer dimerization induced by the nearest-layer hopping $t_1$, which hybridizes the remaining sublattice degrees of freedom into high-energy bonding and antibonding states, thereby pushing them away from the charge neutrality point \cite{zhang2010band, PhysRevB.93.075437}.
A displacement field tends to drive electrons preferentially to either the top or bottom layer depending on its sign.




A key prediction of these effective models is the characteristic power-law dispersion $E \sim |\bm{k}|^N$, where $\bm{k}$ is measured from the $\bm{K}_\tau$ valley \cite{manes2007existence, min2008chiral, koshino2009trigonal}.
This highly nonlinear dispersion gives rise to intrinsically flat low-energy bands, with the degree of flatness increasing rapidly with the number of layers \cite{koshino2010interlayer, slizovskiy2019films}.
Figure \ref{fig:lattice_and_bands}(c). shows the continuum dispersion for one valley of isolated R5G for increasing displacement field energies \cite{dong2024theory}. 
For zero displacement field, there is a band touching at the $\Gamma$ point. 
As the displacement field increased, a band gap develops in conjunction with a flattening of the conduction band.
When subjected to an aligned moiré potential, the continuum band reconstructs into the moiré Brillouin zone.
The moiré $K$ point is at $\approx 0.1 /a_G$, this corresponds to the low-energy conduction moiré bands being formed predominantly by the ``flat'' region of the conduction continuum band.
The resulting suppression of kinetic energy leads to a pronounced enhancement of the density of states (DOS) at the Fermi level, which in turn strongly amplifies electron-electron interactions, as evidenced by the discovery \cite{zhou2021half, zhou2021superconductivity, han2024correlated, liu2023interaction}, in the past few years.
As a consequence, RMG provides a fertile platform for the emergence of correlated quantum phases.

\subsubsection{`Three-patch model'}
In addition, it is worthwhile to mention the `three-patch model' proposed by \cite{soejima2024anomalous}, which offers a markedly different perspective. This approach is explicitly phenomenological, simplifying the Brillouin zone into key patches and focusing on quantum geometry and interaction energies, rather than on the individual hopping parameters characteristic of the SWMC model.

\subsection{Moiré superlattice and Potential Models}

The intrinsic properties of RMG can be further engineered and enhanced through the formation of moiré superlattices. 
Such superlattices typically arise when RMG is aligned with hBN \cite{kennes2021moire}, with a small relative twist angle $\theta$ and a lattice constant mismatch $\epsilon=\frac{a_{hBN}}{a_G} - 1 \approx  1.7\%$. 
The resulting long-wavelength moiré pattern generates a substrate-induced periodic potential $V_{\mathrm{hBN}}$ that strongly modifies the low-energy electronic structure of graphene-based systems.

\subsubsection{superlattice}
\begin{figure}
 \centering
        \includegraphics[width=0.38\textwidth]{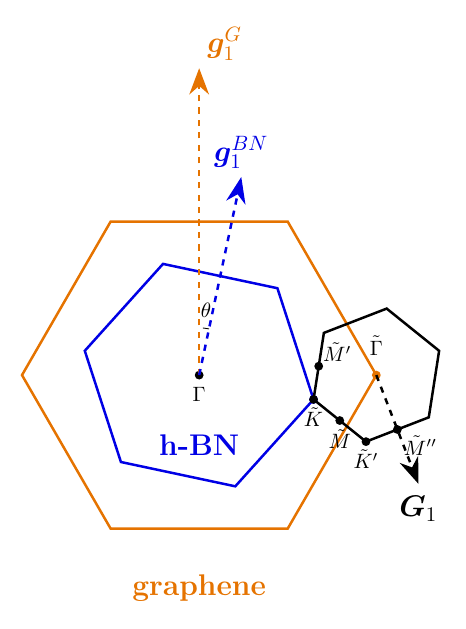}
 \caption{Reciprocal space representation of a graphene/hBN heterostructure. The diagram illustrates the Brillouin zones (BZs) of graphene (large orange hexagon) and hBN (blue hexagon), which are misaligned by a twist angle $\theta$. The vectors $g_1^G$ and $g_1^{BN}$ represent the primary reciprocal lattice vectors for graphene and hBN, respectively. The resulting moiré Brillouin zone (mBZ) is shown as the small black hexagon, centered at the $\bar{\Gamma}$ point, which coincides with the $K$ vertex of the graphene BZ. Key high-symmetry points within the mBZ, including $\bar{K}$, $\bar{K}'$, and the M-points ($\bar{M}$, $\bar{M}'$, $\bar{M}''$), are labeled to indicate electronic folding. The reciprocal lattice vector of the mBZ is defined as $G_1 = g_1^{BN} - g_1^G$.}
\label{fig:moire_k_space}
\end{figure}

We consider the configuration in which the hBN layer is aligned with the bottom graphene layer and rotated clockwise by a small angle $\theta$, as illustrated in Fig.~\ref{fig:moire_k_space}. 
To describe the resulting moiré superlattice, we first introduce the primitive reciprocal lattice vectors of multilayer graphene,
\begin{equation}\label{}
\begin{aligned}
    \bm{g}^{\text{G}}_1 = \frac{2\pi}{a_G} \left(0, \frac{2}{\sqrt{3}} \right), \text{\quad} \bm{g}^{\text{G}}_i= \mathcal{R}_{\frac{(i-1)\pi}{3}} \bm{g}^{\text{G}}_1, \text{ for } i=2,\dots,6
\end{aligned}
\end{equation}
where $\mathcal{R}_{\phi}$ is the anticlockwise rotation operator by angle $\phi$. 
The corresponding reciprocal lattice vectors of hBN are obtained by rescaling and rotating those of graphene,
\begin{equation}
\bm{g}^{\text{BN}}_i = \frac{a_G}{a_{\text{hBN}}} \mathcal{R}_{(-\theta)} \bm{g}^{\text{G}}_i,,\text{\quad for } i=1,\dots,6.
\end{equation}
and the reciprocal lattice vectors of the moiré superlattice are defined as the difference
\begin{equation}
\bm{G}_i = \bm{g}^{\text{BN}}_i - \bm{g}^{\text{G}}_i,\text{\quad for } i=1,\dots,6.
\end{equation}
The magnitude of $\bm{G}_i$ determines the real-space moiré superlattice period (also known as the moiré wavelength) \cite{yankowitz2012emergence, moon2014electronic},
\begin{equation}\label{}
\begin{aligned}
   \lambda = \frac{4\pi}{\sqrt{3}\left|\bm{G}_i\right|} = \frac{ (1+\epsilon)a_G }{\sqrt{ \epsilon^2 + 2(1+\epsilon)(1 - \cos\theta) }}
    \approx \frac{a_{hBN}}{\sqrt{\epsilon^2 + \theta ^2}}
\end{aligned}
\end{equation}
This quantity is directly measurable in experiments and is commonly used to characterize the twist angle and alignment configuration.
For small twist angles $\theta \lesssim 1^\circ$, the corresponding moiré superlattice period typically lies in the range of $10$-$14,\text{nm}$, which is approximately $40$-$60$ times larger than the graphene lattice constant. The moiré pattern is generally rotated with respect to the graphene lattice, and the relative rotation angle can be obtained from the orientation of the moiré reciprocal lattice vectors $\bm{G}_i$ with respect to those of graphene \cite{yankowitz2012emergence}.

\subsubsection{alignment}

While monolayer graphene has $C_6$ rotational symmetry, RMG and hBN both possess only $C_3$ symmetry, with hBN additionally breaking inversion symmetry. 
Because the hBN unit cell contains two inequivalent atoms, boron and nitrogen, and the adjacent graphene layer consists of inequivalent dimer and non-dimer carbon sites, the $C_2$ symmetry that would otherwise relate different stacking configurations is broken, leading to two microscopically distinct alignment configurations of graphene on hBN.
One is the boron atom aligns with the non-dimer carbon atom and the nitrogen atom aligns with the dimer carbon atom, which is regarded as the more ``baseline" alignment pattern and is used in \cite{dong2024theory}; 
the other one is swapped, with the boron atom aligning with the dimer carbon atom and the nitrogen atom aligning with the non-dimer carbon atom. 
In \cite{park2023topological}, these two alignments are referred to as $0^\circ$ and $60^\circ$ orientations of the hBN substrate, denoted as $\xi=1$ and $-1$, respectively. 
(While in \cite{herzog2024moire}, they are denoted as $\xi = 0$ and $1$, respectively. )
This difference arises from the distinct expressions used for their moiré potential.

The two low-energy valleys $\pm K$ of the rhombohedral graphene become decoupled. 
In this limit, valley becomes a good quantum number and we can build separate models for the $K$ and $-K$ valley bands. \cite{herzog2024moire}
We only discuss $H_{K,\xi}(\bm{r})$, the Hamiltonian for the $K$ valley states  since $H_{-K,\xi}(\bm{r}) = H^{*}_{K,\xi}(\bm{r})$, $H_{-K,\xi}(-\bm{k}) = H^{*}_{K,\xi}(\bm{k})$ is obtained from the spinless time-reversal symmetry of graphene.

\subsubsection{moiré potential} When the twist angle between hBN and RMG is small, hereafter referred to as ``aligned'', hBN induces a moiré potential in the graphene layer closest to it. 
The explicit mathematical form of the moiré superlattice potential, which acts only on the first layer adjacent to the aligned hBN, can be derived via different approaches. Here we take the ``baseline'' conﬁguration for example. 

One of them is the tight-binding model \cite{moon2014electronic}, which is applied in \cite{dong2024anomalous, guo2024fractional, herzog2024moire, kwan2025moire}, typically given in the form
\begin{equation}
\begin{aligned}
V_{\mathrm{hBN}}(\bm{r}) = & V_0 \begin{pmatrix} 1 & 0 \\ 0 & 1 \end{pmatrix} + V_1 \left\{ e^{-i \psi} \left[ 
   e^{i \bm{G}_1 \cdot \bm{r}} \begin{pmatrix} 1 & 1 \\ \omega & \omega \end{pmatrix} 
   + e^{i \bm{G}_3 \cdot \bm{r}} \begin{pmatrix} 1 & \omega^2 \\ \omega^2 & \omega \end{pmatrix} 
   + e^{i \bm{G}_5 \cdot \bm{r}} \begin{pmatrix} 1 & \omega \\ 1 & \omega \end{pmatrix} \right] + \text{H.c.} \right\}
\end{aligned}
\end{equation}
where $V_0=28.9\, \mathrm{meV}, V_1=21\, \mathrm{meV}, \psi=-0.29\, \mathrm{rad}$. Its detailed derivation is provided in \cite{moon2014electronic, guo2024fractional, herzog2024moire}.

Another approach is the \textit{ab initio} (DFT) method \cite{jung2015origin, jung2014ab}, which is applied in \cite{chittari2019gate, gonzalez2021topological, patri2023strong, park2023topological, dong2024theory, zhou2024fractional}. It is represented, in the reciprocal space, as following:
\begin{equation}
V_{\mathrm{hBN}}(\bm{k}) = \sum_{\bm{G}} \psi^{\dagger}_{\bm{k} + \bm{G}} \left\{ H_0(\bm{G}) I + H_z(\bm{G}) \sigma_z + \text{Re}[H_{AB}(\mathbf{G})] \sigma_x  + \text{Im}[H_{AB}(\mathbf{G})] \sigma_y \right\} \psi_{\bm{k}}   
\end{equation}
where the diagonal items
\begin{equation}
\begin{aligned}
H_a(\bm{G}_1) &= H_a(\bm{G}_3) = H_a(\bm{G}_5) = C_a e^{i \varphi_a}, \\
H_a(\bm{G}_2) &= H_a(\bm{G}_4) = H_a(\bm{G}_6) = C_a e^{-i \varphi_a}, 
\end{aligned} 
\end{equation}
where $a = 0, z$, $C_0 = -10.13\, \mathrm{meV}, \varphi_0 = 86.53^\circ$ and $C_z = -9.01\, \mathrm{meV}, \varphi_z = 8.43^\circ$. For off-diagonal items $H_{AB}$, 
\begin{equation}
\begin{aligned}
H_{AB}(\bm{G}_1) &= H_{AB}^*(\bm{G}_4) = C_{AB} e^{i \left( \frac{2\pi}{3} - \varphi_{AB} \right)}, \\
H_{AB}(\bm{G}_3) &= H_{AB}^*(\bm{G}_2) = C_{AB} e^{-i \varphi_{AB}}, \\
H_{AB}(\bm{G}_5) &= H_{AB}^*(\bm{G}_6) = C_{AB} e^{i \left( -\frac{2\pi}{3} - \varphi_{AB} \right)}, 
\end{aligned} \tag{B6}
\end{equation}
where $C_{AB} = 11.34\, \mathrm{meV}$ and $\varphi_{AB} = 19.60^\circ$. 

The aforementioned approaches rely on rigid lattices. In addition, several studies have incorporating lattice relaxation \cite{krisna2023moire, jung2017moire}.

\subsection{Hartree-Fock and Mean-Field Approaches for Correlated States}

Hartree-Fock (HF) and self-consistent Hartree-Fock (SCHF) calculations are widely used to approximate the ground state and excited states of these interacting electron systems by treating the many-electron problem as an effective single-particle problem in a mean field. The many-body Hamiltonian, including Coulomb interaction terms, is decoupled into one-body terms and a residual interaction.
In performing the SCHF calculations, the conduction bands of interest are considered, namely the ``active” conduction band, and a number of remote conduction bands are selected within an energy window. 
While the occupied valence bands below the charge neutrality point are neglected as a uniform background for the large displacement fields (interlayer potential difference) \cite{dong2024theory}.

\subsubsection{Coulomb interaction} 
Given the enhanced electron-electron interactions arising from the flat bands in RMG moiré superlattices, many-body theoretical tools become indispensable for understanding the emergent correlated quantum phases.
The high electron density in the flat band suggests the existence of potentially strong interactions. For this reason, we introduce the interaction Hamiltonian to low-energy electrons as
\begin{equation}
\begin{aligned}
    H = \frac{1}{2} \iint d\bm{r}d\bm{r^\prime} :\rho(\bm{r})V(\bm{r}-\bm{r^\prime}) \rho(\bm{r^\prime}) :
\end{aligned}
\end{equation}
where $\hat{\rho}(\bm{r})= \sum_{s=\uparrow ,\downarrow } c^{\dagger}_s(\bm{r}) c_s(\bm{r})$ is the electron density operator at position $\bm{r}$ and $V(\bm{r} - \bm{r'})$ is the interaction potential. After Fourier transformation, it reads
\begin{equation}\label{eq:H_int}
\begin{aligned}
\hat{H}_{\text{int}} &= \frac{1}{2\mathcal{V}} \sum_{\bm{q}} V(\bm{q}) :\hat{\rho}_{-\bm{q}} \hat{\rho}_{\bm{q}}: \\
\end{aligned}
\end{equation}
where $\hat{\rho}_{\bm{q}} = \sum_{\bm{k},s} \hat{c}_{\bm{k},s}^\dagger \hat{c}_{\bm{k} + \bm{q},s} $,
the inverse sample area ${\mathcal{V}}^{-1}= \frac{A_{\bm{k}}}{(2\pi)^2}$, $A_{\bm{k}}$ is the area of the discrete grid in $\bm{k}$-space. $:\hat{\mathcal{O}}:$ donates the normal ordering of operator $\hat{\mathcal{O}}$.

To mimic the experiment in \cite{lu2024fractional}, we apply a dual gate-screened Coulomb interaction \cite{yang2025two} to the low-energy electrons with the screened potential $V(\bm{q})=\frac{e^2 }{2\epsilon_0 \epsilon_r q} \tanh (qd_s)$, where $e$ is the unit charge, $\epsilon_0$ is the vacuum permittivity, $\epsilon_r$ is the relative permittivity, $d_s$ is the screening length ,i.e., the distance between the the double screening metallic gates.

It should be noted that the SI unit system is adopted in the present work. Some studies\cite{zhou2021half,dong2024anomalous}, however, employ the Gaussian unit system, which differs from the SI system by a factor of $4\pi\epsilon_0$; consequently, the corresponding potential takes the form $V^{\text{Gauss}}_{\bm{q}}=\frac{2\pi e^2 }{\epsilon_r q} \tanh (qd_s)$.
\cite{zhou2024fractional} takes the $\hat{z}-$direction (layer) into account and brings in a factor $e^{-q|l-l^\prime|c}$, where $c$ is the aforementioned inter-layer distance, $l$($l^\prime$) is the layer indices of the interaction electrons, which indicates the interaction strength exhibits an exponential decay with the number of spacer layers, with the electrons preserving (or exchanging) their layer indices before and after the interaction.
The single gate-screened Coulomb interaction $V(\bm{q})=\frac{e^2 }{2\epsilon_0 \epsilon_r q} (1-e^{-2qd_s})$ is employed in \cite{zhang2019bridging},  as used in twisted bilayer graphene \cite{bultinck2020ground, liu2021nematic, zhang2020correlated}.
The Yukawa potential $V(\bm{q})=\frac{e^2 }{2 \epsilon_0 \epsilon_r } \frac{1}{\sqrt{q^2+\kappa^2}}$ is employed in \cite{abouelkomsan2020particle}. 
As a similar idea, \cite{zhang2019nearly} uses the Coulomb potential $V(\bm{q})\sim \frac{1}{q}$ as a toy model which corresponds to  $V(\bm{r})\sim \frac{1}{r}$ in 2D real space.
In short, the main features of the physical results are robust against 
 different forms of interaction used
in the calculations.

\subsubsection{Second quantization form of interaction}

The the Bloch state calculated from the continuum model has the following form
\begin{equation}\label{}  
    \hat{\psi}^{\dagger}_{\bm{k} s \tau  n} = \sum_{\alpha, \bm{G}} u_{\tau, n,\bm{G}, \alpha} (\bm{k}) \hat{c}^{\dagger}_{\alpha,\bm{k}+\bm{G}, s \tau} 
\end{equation}
with inverse transformation
\begin{equation}\label{eq:psi_to_c}  
  \hat{c}^{\dagger}_{\alpha,\bm{k}+\bm{G}, s \tau}  =  \sum_{n} u^{*}_{\tau, n,\bm{G}, \alpha} (\bm{k}) \hat{\psi}^{\dagger}_{\bm{k} s \tau  n}
\end{equation}
where the momentum of the operator $\hat{\psi}$ is defined in the moiré
Brillouin Zone (mBZ), with the periodic boundary condition $\hat{\psi}_{\bm{k} s \tau  n} = \hat{\psi}_{(\bm{k}+\bm{G}) s \tau  n}$ for $\bm{G}$ the reciprocal lattice of moiré.
Here we recall that $s\in\{\uparrow, \downarrow\}$, $\tau \in \{+,-\} $, $\alpha \in \{A_1, \dots, B_N \}$, $n$ are spin, valley, sublattice and band indices, respectively. 
Besides, we take the valley in consideration by replacing $\hat{\rho}_{\bm{q}}$ in Eq.\,\ref{eq:H_int} with 
\begin{equation}
\begin{aligned}
  \hat{\rho}_{\bm{q}+\bm{K}_{\tau^\prime}-\bm{K}_{\tau}} &=\sum_{\alpha, \bm{k},s } \sum_{\tau, \tau^\prime} \hat{c}_{\alpha, \bm{k},s \tau}^\dagger \hat{c}_{\alpha, \bm{k} + \bm{q},s \tau^\prime}
  =\sum_{\tau} (\hat{\rho}_{\bm{q}, \tau}^{+} + \hat{\rho}_{\bm{q}, \tau}^{-}) \\
\end{aligned}
\end{equation}
where $
  \hat{\rho}_{\bm{q}, \tau}^{\pm} = \sum_{\alpha, \bm{k},s } \hat{c}_{\alpha, \bm{k},s \tau}^\dagger \hat{c}_{\alpha, \bm{k} + \bm{q},s,\pm\tau} $
are the intravaley($+$) and intervalley($-$) density operators, respectively.
$\hat{c}_{\alpha,\bm{k}, s \tau }$ is the electron operator at momentum space point $K_{\tau}+\bm{k}$ sublattice $\alpha$ with spin $s$.
Define $\tilde{\bm{k}} = \bm{k} + \bm{G}$, $\tilde{\bm{q}} = \bm{q} + \bm{Q}$ where $\tilde{\bm{k}}, \tilde{\bm{q}}\in \text{BZ}$, ${\bm{k}}, {\bm{q}}\in \text{mBZ}$, $\bm{G}$ and $\bm{Q}$ are the reciprocal lattice of moiré, and substituting Eq.\,\ref{eq:psi_to_c} then
\begin{equation}\label{eq:rho_plus}
\begin{aligned}
  \hat{\rho}_{\tilde{\bm{q}}, \tau}^{\pm} &= 
  \sum_{\alpha, \tilde{\bm{k}}, \tilde{\bm{k}}^\prime,s } \hat{c}_{\alpha, \tilde{\bm{k}},s \tau}^\dagger \hat{c}_{\alpha, \tilde{\bm{k}}^\prime,s,\pm\tau} \delta_{\tilde{\bm{k}} + \tilde{\bm{q}}, \tilde{\bm{k}}^\prime}\\
  &= \sum_{\alpha, \bm{k}, \bm{k}^\prime, \bm{G}, \bm{G}^\prime, s }
  \hat{c}_{\alpha, \bm{k} + \bm{G},s \tau}^\dagger 
  \hat{c}_{\alpha, \bm{k}^\prime + \bm{G}^\prime,s,\pm\tau} 
  \delta_{\bm{k}+ \bm{G} + \tilde{\bm{q}}, \bm{k}^\prime+ \bm{G}^\prime } \\
  &= \sum_{\alpha, \bm{k}, \bm{k}^\prime, \bm{G}, \bm{G}^\prime, s }
  \sum_{n} u^{*}_{\tau, n,\bm{G}, \alpha} (\bm{k}) \hat{\psi}^{\dagger}_{\bm{k} s \tau  n}
  \sum_{n^\prime} u^{}_{\pm\tau, n^\prime,\bm{G}^\prime, \alpha} (\bm{k}^\prime) \hat{\psi}_{\bm{k}^\prime s,\pm \tau  n^\prime}
  \delta_{\bm{k}+ \bm{G} + \tilde{\bm{q}}, \bm{k}^\prime+ \bm{G}^\prime } \\
  &= \sum_{\bm{k}, \bm{k}^\prime,  s , n,n^\prime }
  \Lambda^{\tau, \pm}_{\tilde{\bm{q}}, \bm{k}, \bm{k}^\prime; n, n^\prime}
     \hat{\psi}^{\dagger}_{\bm{k} s \tau  n}
     \hat{\psi}_{\bm{k}^\prime s,\pm \tau  n^\prime}   \\
\end{aligned}
\end{equation}
where
\begin{equation}
\begin{aligned}
  \Lambda^{\tau, \pm}_{\tilde{\bm{q}}, \bm{k}, \bm{k}^\prime; n, n^\prime} &= 
  \sum_{\alpha, \bm{G}, \bm{G}^\prime} u^{*}_{\tau, n,\bm{G}, \alpha} (\bm{k}) u^{}_{\pm\tau, n^\prime,\bm{G}^\prime, \alpha} (\bm{k}^\prime) 
  \delta_{\bm{k}+ \bm{G} + \tilde{\bm{q}}, \bm{k}^\prime+ \bm{G}^\prime } \\
\end{aligned}
\end{equation}
Note that in real numerical calculations, $-\bm{q}$ may not in the selected mBZ, so it is helpful to transform $\tilde{\bm{k}}\leftrightarrow \tilde{\bm{k}}^\prime$, ${\tilde{\bm{q}}}\leftrightarrow -{\tilde{\bm{q}}}$, then
\begin{equation}
\begin{aligned}
  \left( \Lambda^{\tau, \pm}_{-\tilde{\bm{q}}, \bm{k}^\prime, \bm{k} ; n^\prime, n} \right)^{*} &= 
  \sum_{\alpha, \bm{G}, \bm{G}^\prime} u_{\tau, n^\prime,\bm{G}^\prime, \alpha} (\bm{k}^\prime) u^{*}_{\pm\tau, n,\bm{G} , \alpha} (\bm{k} ) 
  \delta_{\bm{k}+ \bm{G} - \tilde{\bm{q}}, \bm{k}^\prime+ \bm{G}^\prime }
  = \Lambda^{\tau, \pm}_{\tilde{\bm{q}}, \bm{k}, \bm{k}^\prime; n, n^\prime}  \\
\end{aligned}
\end{equation}

At this point, we split the interaction Hamiltonian Eq.\,\ref{eq:H_int} into intravalley and intervalley parts as
$ \hat{H}_{\text{int}} = \hat{H}_{\text{intra}} + \hat{H}_{\text{inter}} $ where
\begin{equation}\label{eq:h_intra}
\begin{aligned}
   \hat{H}_{\text{intra}} &= \frac{1}{2\mathcal{V}} \sum_{\tau, \tau^\prime} \sum_{\tilde{\bm{q}}} V(\tilde{\bm{q}}) : \hat{\rho}_{-\tilde{\bm{q}}, \tau^\prime}^{+}  \hat{\rho}_{\tilde{\bm{q}}, \tau}^{+}: \\
\end{aligned}
\end{equation}

\begin{equation}
\begin{aligned}
   \hat{H}_{\text{inter}} &= \frac{1}{2\mathcal{V}} \sum_{\tau} \sum_{\tilde{\bm{q}}} V(\tilde{\bm{q}}+\bm{K}_{-\tau}-\bm{K}_{\tau}) : \hat{\rho}_{-\tilde{\bm{q}}, -\tau}^{-}  \hat{\rho}_{\tilde{\bm{q}}, \tau}^{-}: \\
\end{aligned}
\end{equation}

As explained in \cite{liu2021theories,lee2019theory}  If valley symmetry is exact, states belonging to different valleys would be orthogonal leading to a vanishing intervalley density $\rho_{\bm{q},\tau}^{-}$.
However, if valley symmetry is broken on the scale of $V(\bm{K}_{+} - \bm{K}_{-})$, an effective intervalley Hund's coupling term emerges. This term is usually neglected since it is much smaller than the interaction between intravalley densities. 
Nevertheless, contributions from this term can lift the degeneracy between
different broken symmetry states which are otherwise exactly degenerate, which makes it important to include it in some cases.

\subsubsection{Projection}

Following the approach in \cite{lee2019theory, liu2021theories,zhang2020correlated, chatterjee2022inter}, we  perform  Hartree-Fock approximation by keeping the first $N_b$ conduction bands in the moiré Brillouin zone (mBZ). 
To this end, the reference state in the normal ordering is chosen to be the charge-neutrality scheme, i.e. all valence bands are full and all conduction bands are empty. 
It should be noted that the orthogonality relation for the Bloch states $|\hat{\psi}^{\dagger}_{\bm{k} s \tau n} \rangle$ is satisfied naturally, i.e.,
\begin{equation}\label{}  
\langle \hat{\psi}_{\bm{k} s \tau  n^\prime} | \hat{\psi}_{\bm{k} s \tau  n} \rangle = \sum_{\alpha, \bm{G}} u^*_{\tau, n^\prime,\bm{G}, \alpha} (\mathbf{k}) u_{\tau, n,\bm{G}, \alpha} (\mathbf{k}) = \delta_{n, n^\prime}
\end{equation}
But the completeness relation is no longer strictly satisfied, i.e.,
\begin{equation}\label{}  
\sum_{n =1}^{N_b} | \hat{\psi}_{\bm{k} s \tau  n} \rangle \langle \hat{\psi}_{\bm{k} s \tau  n} | = \sum_{n =1}^{N_b} \sum_{\alpha, \alpha^\prime } \sum_{\bm{G}, \bm{G}^\prime} u_{\tau, n,\bm{G}, \alpha} (\mathbf{k}) u^*_{\tau, n,\bm{G}^\prime, \alpha^\prime} (\mathbf{k}) | c_{\alpha,\bm{k}+\bm{G}, s \tau} \rangle \langle  c_{\alpha^\prime,\bm{k}+\bm{G}^\prime, s \tau}  | \neq \mathbbm{1}
\end{equation}
since the summation of the band is limitted in the low-energy ones rather than all bands, and this is the origin of the term ``projection''---we are projecting the full Hilbert space onto a low-energy active subspace spanned by the $\hat{\psi}^{\dagger}_{\bm{k} s \tau  n}$ operators.

Besides the charge-neutrality scheme, it also worth to study \cite{kwan2025moire, yu2025moire} the average scheme where the density operator in real and momentum spaces are replaced with
\begin{equation}
\begin{aligned}
  \hat{\rho}(\bm{r}) \to \delta \hat{\rho}(\bm{r})
  = \sum_{s } \left[ c^{\dagger}_s(\bm{r}) c_s(\bm{r}) - \frac{1}{2}\delta(\bm{r})\right]
  = \sum_{s } \left[ c^{\dagger}_s(\bm{r}),  c_s(\bm{r}) \right] \\
  \hat{\rho}_{\bm{q}} \to \delta\hat{\rho}_{\bm{q}} 
  = \sum_{\bm{k},s} \left[ \hat{c}_{\bm{k},s}^\dagger \hat{c}_{\bm{k} + \bm{q},s} - \frac{1}{2}\{ \hat{c}_{\bm{k},s}^\dagger,  \hat{c}_{\bm{k} + \bm{q},s}  \}\right] 
  = \sum_{\bm{k},s} \left[ \hat{c}_{\bm{k},s}^\dagger, \hat{c}_{\bm{k} + \bm{q},s} \right]
\end{aligned}
\end{equation}
respectively.

\subsubsection{ Hartree-Fock approximation}

Substituting Eq.\,\ref{eq:rho_plus} to the intravalley interaction Hamiltonian Eq.\,\ref{eq:h_intra} as:
\begin{equation}\label{eq:intra}
\begin{aligned}
   \hat{H}_{\text{intra}} &= \frac{1}{2\mathcal{V}} \sum_{\tau_1, \tau_2} \sum_{\tilde{\bm{q}}} V(\tilde{\bm{q}}) : \tilde{\rho}_{-\tilde{\bm{q}}, \tau_1}^{+}  \tilde{\rho}_{\tilde{\bm{q}}, \tau_2}^{+}: \\
   &= \frac{1}{2\mathcal{V}} \sum_{\tau_1, \tau_2, s_1 ,s_2}  
   \sum_{\{ \bm{k}_i\} } \sum_{ \{n_i \} }
   V^{\tau_2 \tau_1}_{n_4 n_3 n_2 n_1}(\bm{k}_1,\bm{k}_2,\bm{k}_3, \bm{k}_4)
     \hat{\psi}^{\dagger}_{\bm{k}_4 s_1 \tau_1  n_4}
     \hat{\psi}^{\dagger}_{\bm{k}_3 s_2 \tau_2  n_3}
     \hat{\psi}_{\bm{k}_2 s_2 \tau_2  n_2}  
     \hat{\psi}_{\bm{k}_1 s_1 \tau_1  n_1}  \\
\end{aligned}
\end{equation}
where 
\begin{equation}\label{}
\begin{aligned}
  \hat{\rho}_{-\tilde{\bm{q}}, \tau_1}^{+} 
  &= \sum_{s_1 ,\bm{k}_4, \bm{k}_1,   n_4,n_1 }
  \left( \Lambda^{\tau_1, +}_{\tilde{\bm{q}}, \bm{k}_1, \bm{k}_4; n_1, n_4} \right)^{*}
     \hat{\psi}^{\dagger}_{\bm{k}_4 s_1 \tau_1  n_4}
     \hat{\psi}_{\bm{k}_1 s_1 \tau_1  n_1}   \\
  \hat{\rho}_{\tilde{\bm{q}}, \tau_2}^{+} 
  &= \sum_{s_2 ,\bm{k}_2, \bm{k}_3,   n_2,,n_3 }
  \Lambda^{\tau_2, +}_{\tilde{\bm{q}}, \bm{k}_3, \bm{k}_2; n_3, n_2}
     \hat{\psi}^{\dagger}_{\bm{k}_3 s_2 \tau_2  n_3}
     \hat{\psi}_{\bm{k}_2 s_2 \tau_2  n_2}   \\
\end{aligned}
\end{equation}
and 
\begin{equation}
\begin{aligned}
   V^{\tau_2 \tau_1}_{n_4 n_3 n_2 n_1}(\bm{k}_1,\bm{k}_2,\bm{k}_3, \bm{k}_4) &:= \sum_{\bm{Q}} V(\bm{q}+\bm{Q})  \Lambda^{\tau_2, +}_{\bm{q}+\bm{Q}, \bm{k}_3, \bm{k}_2; n_3, n_2}
  \left( \Lambda^{\tau_1, +}_{\bm{q}+\bm{Q}, \bm{k}_1, \bm{k}_4; n_1, n_4} \right)^{*}
\end{aligned}
\end{equation}
In addition, we define the order parameter as $P(\bm{k}, \bm{k}^\prime)_{s\tau n, s^\prime \tau^\prime n^\prime} = \langle \hat{\psi}_{\bm{k}s\tau n}^\dagger \hat{\psi}_{ \bm{k}^\prime s^\prime \tau^\prime n^\prime} \rangle$ correspondence with Slater determinant states. Because the continuum Hamiltonian  preserves the translational symmetry of the moiré superlattice, states in different momentum sectors cannot mix, and the order parameter is diagonal as $P(\bm{k})_{s\tau n, s^\prime \tau^\prime n^\prime} = \delta_{\bm{k}, \bm{k}^\prime} P(\bm{k}, \bm{k}^\prime)_{s\tau n, s^\prime \tau^\prime n^\prime}$.
In the absence of spin-orbit coupling or an external magnetic field, the spin is conserved. Thus, one can restrict to $S_z$-conserving states and further assume that $P(\bm{k})$ is diagonal in the spin space.
In the absence of intervalley interaction like $\hat{H}_{\text{inter}}$, the valley is also conserved and two valleys are connected by $SU(2)$ symmetry. Thus, one can further assume that $P(\bm{k})$ is also diagonal in the valley space.

For convenience, denote
\begin{equation}\label{eq:V_symmetry}
\begin{aligned}
   V^{\tau_2 \tau_1, \text{H}}_{n_4 n_3 n_2 n_1}(\bm{k}_1,\bm{k}_2) &:= V^{\tau_2 \tau_1}_{n_4 n_3 n_2 n_1}(\bm{k}_1,\bm{k}_2,\bm{k}_2, \bm{k}_1) \\
   V^{\tau_2 \tau_1, \text{F}}_{n_4 n_3 n_2 n_1}(\bm{k}_1,\bm{k}_2) &:= V^{\tau_2 \tau_1}_{n_4 n_3 n_2 n_1}(\bm{k}_1,\bm{k}_2,\bm{k}_1, \bm{k}_2)
\end{aligned}
\end{equation}
%
With the help of Wick theory (See details in Appendix), the Hartree-Fock approximation is given by
\begin{equation}
\begin{aligned}
   \hat{H}_\text{intra}^{\text{HF}}  
     =& \frac{1}{\mathcal{V}} \sum_{\bm{k}_1} \sum_{\tau_1, \tau_2, s_1 ,s_2}
   \sum_{ n_1, n_2 }
  \hat{\psi}^\dagger_{\bm{k}_1 s_1 \tau_1 n_1} 
  h_{\text{HF}}(\bm{k}_1)_{ s_1 \tau_1 n_1,  s_2 \tau_2 n_2}
  \hat{\psi}_{\bm{k}_1 s_2 \tau_2 n_2}  - E_{\text{intra}}^{\text{C}} \\
\end{aligned}
\end{equation}
where
\begin{equation}
\begin{aligned}
  h_{\text{HF}}(\bm{k}_1)_{ s_1 \tau_1 n_1,  s_2 \tau_2 n_2} &= \sum_{ \bm{k}_2 }  \sum_{ n_3, n_4 } \left[ \delta_{s_1 , s_2} \delta_{\tau_1, \tau_2} \sum_{\tau_3, s_3} V^{\tau_3 \tau_1, \text{H}}_{n_1 n_3 n_4 n_2}(\bm{k}_1,\bm{k}_2) P(\bm{k}_2)_{s_3 \tau_3 n_3, s_3 \tau_3 n_4} \right.  \\
  & \text{\qquad\qquad\qquad\qquad}  \left.
   - V^{\tau_1 \tau_2, \text{F}}_{n_3 n_1 n_4 n_2}(\bm{k}_1,\bm{k}_2) P(\bm{k}_2)_{s_2 \tau_2  n_3, s_1 \tau_1 n_4}   \right] \\
\end{aligned}
\end{equation}
and the condensed energy
\begin{equation}
\begin{aligned}
E_{\text{intra}}^{\text{C}}
&= \frac{1}{2\mathcal{V}} \sum_{\tau_1, \tau_2, s_1 ,s_2}
\sum_{\{ \bm{k}_i\} } \sum_{ \{n_i \} }
\left[ V^{\tau_2 \tau_1, \text{H}}_{n_4 n_3 n_2 n_1}(\bm{k}_1,\bm{k}_2) P(\bm{k}_1)_{s_1 \tau_1 n_4, s_1 \tau_1 n_1 } P(\bm{k}_2)_{s_2 \tau_2 n_3, s_2 \tau_2 n_2 }
 \right. \\
&  \text{\qquad\qquad\qquad\qquad\qquad\quad} \left.  - V^{\tau_2 \tau_1, \text{F}}_{n_4 n_3 n_2 n_1}(\bm{k}_1,\bm{k}_2) P(\bm{k}_2)_{s_1 \tau_1  n_4, s_2 \tau_2 n_2}  P(\bm{k}_1)_{s_2 \tau_2  n_3, s_1 \tau_1  n_1} \right] \\
\end{aligned}
\end{equation}

\subsection{Chern Numbers and QAH Effect}

The Berry curvature itself is a geometrical property of the Bloch wave functions in momentum space, defined as 
$\Omega(\bm{k}) = i \left[ \langle \partial_{k_x} \psi_{\bm{k}} | \partial_{k_y} \psi_{\bm{k}} \rangle -  \langle \partial_{k_y} \psi_{\bm{k}} | \partial_{k_x} \psi_{\bm{k}} \rangle \right]$, where $|\psi_{\bm{k}}\rangle$ are the periodic parts of the Bloch wave functions.
The Chern number, a topological invariant, is achieved through the integration of the Berry curvature over the moiré Brillouin zone as
$C = \frac{1}{2\pi} \int_{BZ}  \Omega(\bm{k})  \,\mathrm{d}^2\bm{k}$. 
The detailed numerical algorithm \cite{fukui2005chern} for calculating the Chern number is shown in Appendix.
A non-zero integrated Berry curvature signifies a topologically non-trivial band. 

The QAH effect is a captivating topological phenomenon where a material exhibits quantized Hall conductance $\sigma_{xy}$ in the absence of an external magnetic field. 
The Hall conductance is directly proportional to this integer Chern number, 
and the Streda formula \cite{streda1982theory}, which relates the Hall conductivity  to the rate of change of charge density $n$ with respect to the magnetic field $B$, 
$\sigma_{xy} = C \frac{e^2}{h} = \frac{\partial n}{\partial B}$,
provides an experimental means to infer the Chern number from transport measurements.

The realization of the QAH effect often relies on spontaneously breaking time-reversal symmetry of the ground state, typically through intrinsic ferromagnetism in materials. 

At a filling factor of $\nu=1$ of this interaction-induced band, the system realizes an Integer QAH insulator with $\sigma_{xy} = \pm e^2/h$. 
This is achieved through spontaneous valley and spin polarization, which breaks the time-reversal symmetry.

\section{Theoretical description of QAH effect in RMG/hBN}
\label{sec:theory}

\begin{figure}
 \centering
        \includegraphics[width=0.9\textwidth]{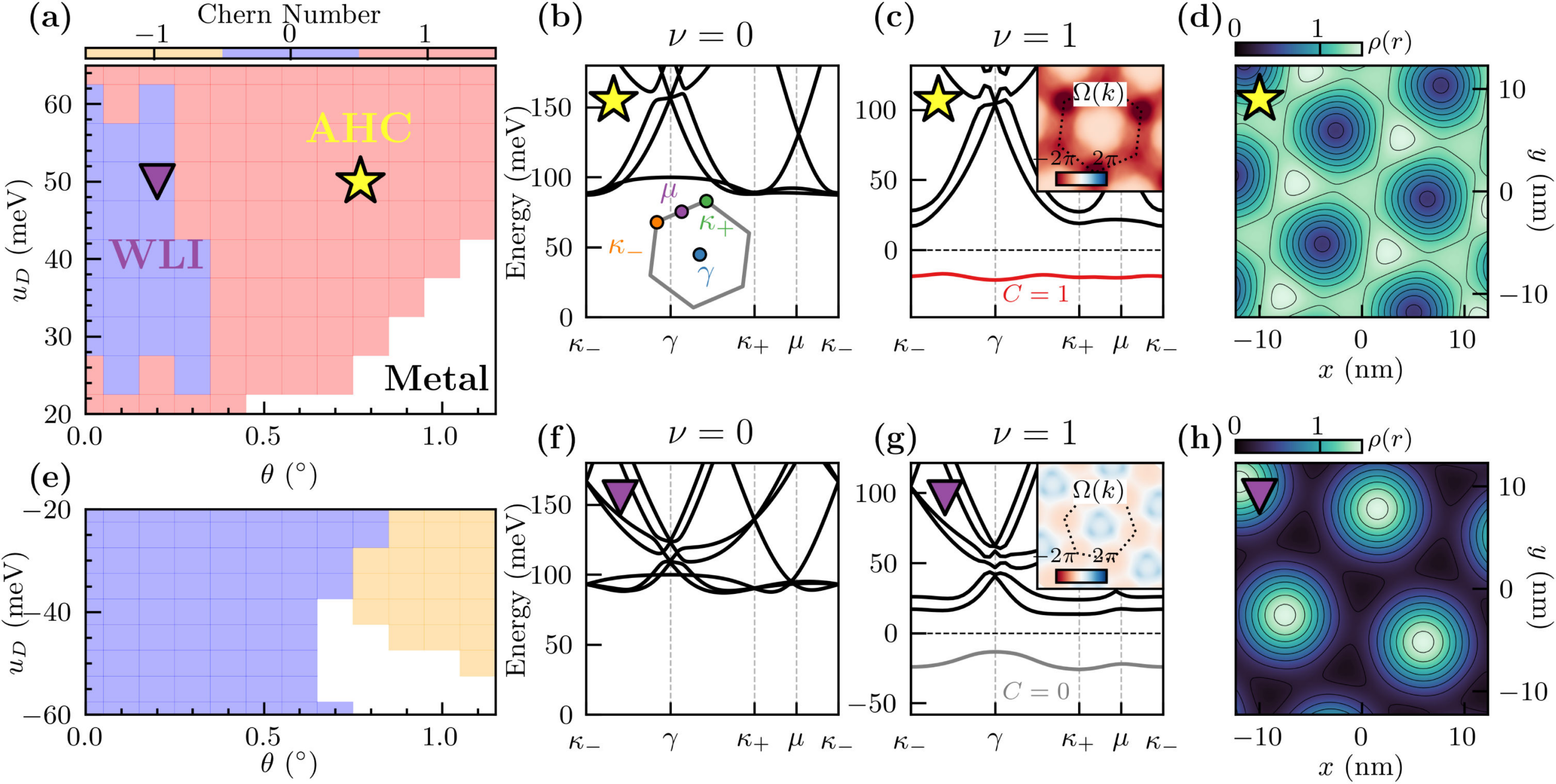}
 \caption{
   (Cited from \cite{dong2024anomalous})  
   (a),(e) The phase diagram of moiré rhombohedral pentalayer graphene as a function of the interlayer potential difference $\Delta$, and the twist angle $\theta$. The colors label the Chern number of the SCHF ground states, with gapless regions shown in white. 
   On the moiré-distant side ($\Delta < 0$), the anomalous Hall crystal (AHC) phase has $C=1$, whereas the Wigner-like insulator (WLI) has $C=0$. 
   (b),(f) Noninteracting band structures at the yellow star ($-\Delta, \theta) =(50\,\mathrm{meV; 0.77^\circ})$ and purple triangle ($-\Delta, \theta) =(50\,\mathrm{meV; 0.2^\circ})$. 
   Inset in (b): plot of the moiré Brillouin zone. 
   (c),(g) SCHF band structure at $\nu=1$, finding an AHC and WLI at the yellow star and purple triangle, respectively. 
   Insets: Berry curvature $\Omega(k)$ of the occupied band. (d),(h) Charge densities $\rho(r)$ of the AHC (d) and WLI (h).  
   The AHC charge density resembles a honeycomb lattice, whereas the WLI charge density resembles a triangular lattice.
   }
\label{fig:Theoretical}
\end{figure}

\subsection{numerical results}

\subsubsection{Chern-1 band}
We start with the integer quantum Hall effect with $\sigma_{xy} = \frac{e^2}{h}$ at $\nu = 1$ of the conduction band observed in the experiments of \cite{lu2024fractional} on pentalayer rhombohedral graphene aligned with hBN, with a small twist angle $\theta = 0.77^\circ$.
The interlayer potential difference per unit charge $\Delta$ is used to simulate the perpendicular external displacement field $D=\epsilon_r E = \epsilon_r \Delta/c$, where $c$ is pre-defined the interlayer distance and $\epsilon_r$ is the relative permittivity also used in Hartree-Fock.
The sign of $\Delta$ is selected such that for the positive $\Delta$, electrons are pushed towards and filled from the higher-index layer \cite{zhou2024layer}, thus the conduction band electrons are mainly localized in the lowest-index layer, i.e., the aforementioned ``moiré-proximate'' side. 
For the negative $\Delta$, the electrons are filled from the lower-index layer, thus the conduction band electrons are mainly localized in the highest-index layer, i.e., the aforementioned ``moiré-distant'' side \cite{dong2024anomalous}.

Figure \,\ref{fig:Theoretical}(b) shows the band structure of the non-interacting Hamiltonian
\begin{equation}
    h_{\text{kin}} = {H}_{N=5} + V_{\mathrm{hBN}}
\end{equation}
at $\theta = 0.77^\circ$ and $\Delta = -50$ meV, where the displacement field opens a gap between the conduction and valence bands.
Thus, it is natural to assume that all valence bands are fully filled at charge neutrality. 
However, no clear and direct energy gap exists around $\nu=1$ in conduction bands. This implies that the observed gapped phase at $\nu=1$ may have an interaction origin. Thus, several different groups \cite{dong2024anomalous,zhou2024fractional,dong2024theory,guo2024fractional} proposed to consider the following the Hamiltonian
\begin{equation}
   \hat{\mathcal{H}} = \hat{h}_{\text{kin}} + H_{\text{int}}
\end{equation}
with electron-electron interactions treated within the self-consistent Hartree-Fock mean-field method.
Besides, the experimental observation of the non-zero Chern band \cite{lu2024fractional} implies that a single valley and spin polarized band is realized at the filling $\nu=1$ \cite{dong2024theory}, which is also supported by the theoretical considerations \cite{zhang2019nearly}. 
Numerical results show that interactions open a gap, where the occupied isolated band (red in Figure\,\ref{fig:Theoretical}(c)) has bandwidth 6 meV and nonzero Chern number $|C| = 1$. 

\subsubsection{Stability of Chern-1 band}
Systematic variations of model parameters consistently demonstrate the robustness of the $C=1$ state in the  ``moiré-distant'' side, which can be understood from three complementary perspectives.

At the most global level, the $C=1$ phase is stable against changes in hyperparameters such as the twist angle $\theta$, interlayer potential difference $\Delta$, and layer number $N$.
Figure~\ref{fig:Theoretical}(a) presents the Chern number as a function of the twist angle $\theta$ and the interlayer potential difference $\Delta$, revealing an extended $C=1$ region on the negative-$\Delta$ side.
Consistent with this observation, Ref. \cite{dong2024anomalous,dong2024theory,zhou2024fractional} identified similar parameter regimes supporting isolated flat bands with $C=1$ for layer numbers $N=3 \sim 7$. 
This agrees with the expectation that at large displacement fields only the top two layers away from the hBN have any occupation, therefore the total number of layers does not matter much. However, the optimal range of the interlayer potential difference $\Delta$ decreases with increasing layer number $N$, as expected from the fact that the bands are flatter for larger $N$ \cite{dong2024theory}.

Additionally, the $C=1$ phase also exhibits pronounced stability with respect to the strength of the moiré potential.
In particular, tuning the relative permittivity $\epsilon_r$, or introducing an scaling factor for the moiré potential~\cite{zhou2024fractional}, modifies the phase boundaries but leaves the Chern-1 phase intact over a broad parameter range.

Finally, this robustness persists even at the microscopic graphene scale.
As demonstrated in \cite{dong2024anomalous}, varying the Fermi velocity (via $t_0$) substantially shifts the phase boundaries without altering the set of realized phases, while tuning the trigonal warping terms ($t_{2,3,4}$ in Eq.~\ref{eq:H_N_block})—or even setting them to zero by retaining only nearest-neighbor hoppings— does not destabilize the $C=1$ phase.
This insensitivity to graphene-scale details indicates that higher-order hopping terms play only a secondary role and highlighting the universal and minimal nature of the $C=1$ phase.

\subsection{Is moiré potential necessary? }

The role of the moiré potential in RMG/hBN remains contentious. 
While Zhou et al. \cite{zhou2021half, zhou2021superconductivity} suggest that in intrinsic rhombohedral trilayer graphene (R3G) electronic correlations drive various flavor-ferromagnetic metallic states and unconventional superconductivity, relegating the moiré potential to a perturbative role.
Patri and Senthil \cite{patri2023strong}, by contrast, argue for the structural importance of the moiré potential in determining the stability of correlated insulating phases.
Furthermore, Galeano González \textit{et al.} \cite{gonzalez2021topological} highlighted that topological phase transitions can be triggered by pseudomagnetic vector field potentials associated with moiré strain patterns. 
Therefore, the specific contribution of the moiré potential to the realization of Chern insulators and the QAH effect in RMG/hBN demands independent and critical investigation, which we pursue here by addressing a more fundamental question: whether the moiré superlattice potential is an essential ingredient for the formation of the $\nu = 1$, $C = 1$ Chern band, whose robustness against variations of model parameters has been established above.

Before discussing this question, we note that on the ``moiré-proximate'' side ($\Delta > 0$) \cite{dong2024anomalous}, the dominant phase is a $C = 0$ insulator (consistent with experiment \cite{lu2024fractional}), which is governed by the details of $V_{\mathrm{hBN}}$, and that varying the moiré potential strength and phase can induce a transition to a metal or even a Chern insulator \cite{gonzalez2021topological}. So the moiré potential is clearly important on this side.

As for the ``moiré-distant'' side ($\Delta < 0$), as it discussed above, the Chern-1 band is robust and persists even in the absence of a moiré potential and trigonal warping terms of graphene. This indicates that interactions alone are sufficient to stabilize the topological band.

From an energy-scale perspective, the moiré potential in RMG is expected to be a weak perturbation \cite{zhou2024fractional}. 
For instance, in the pentalayer rhombohedral graphene aligned with hBN, the projected moiré potential onto the conduction band is estimated to be only on the order of 0.05 meV, which is negligible compared to both the conduction bandwidth and the Coulomb interaction scale. 
Correspondingly, the band structures with and without moiré potential are nearly indistinguishable. 
Additionally, SCHF calculations show that tuning the phase \cite{dong2024anomalous} or reducing the strength \cite{zhou2024fractional} of the moiré potential has only a very weak effect on the magnitude of the band gap.

Taken together, these results indicate that the moiré potential is not required to generate the $\nu=1$ Chern insulator. 
Instead, its primary role is to weakly pin and stabilize an interaction-driven phase that already exists in the absence of a superlattice. 
This observation motivates a physical interpretation of the $C=1$ state as a symmetry-breaking phase driven by Coulomb interactions, which we discuss in detail in the next subsection.

In summary, the $\nu=1$ QAH insulator is fundamentally an interaction-driven phenomenon. The moiré superlattice mainly facilitates some experimental reasons (such as weakly pinning down the electronic orders), while the essential topology and band isolation originate from electron-electron interactions. This perspective complements the previously discussed robustness of the $C=1$ band against microscopic parameters and layer number, emphasizing the universal and minimal nature of the interaction-driven Chern insulator.

\subsection{Anomalous Hall Crystal (AHC)}

In this subsection, we discuss the many-body ground state associated with the Chern-1 band at filling $\nu=1$, which can be naturally interpreted as an AHC. This phase is not directly induced by the moiré potential or by specific microscopic parameters of graphene; instead, it emerges only self-consistently from electron-electron interactions within Hartree-Fock theory. Its defining feature is the simultaneous appearance of nontrivial topological response and spontaneous breaking of continuous translational symmetry, combining topology and crystalline order.

The physical picture of the AHC is closely related to the concept of a ``Hall crystal", originally introduced by Halperin \cite{halperin1986compatibility} and Tešanović \cite{tevsanovic1989hall} to describe states that exhibit quantized Hall response while spontaneously breaking translational symmetry.
It is crucial to distinguish the AHC discussed here from single-particle Chern bands generated by externally imposed superlattice potentials, such as those realized in twisted bilayer transition metal dichalcogenides.
In those systems, the Chern bands are single-particle Bloch bands defined with respect to the moiré Brillouin zone, whereas in AHC picture the Chern-1 band and its associated quantum state arise as an interaction-driven many-body phenomenon.

To further clarify the origin of the AHC, the Hamiltonian without any moiré potential is first considered \cite{dong2024anomalous,zhou2024fractional}.
In this limit, numerical calculations show the system  reduces to pristine RMG, with the low-energy conduction band filled to the same electron density. 
At the noninteracting level, this corresponds to a metallic state that preserves graphene's intrinsic continuous translational symmetry and does not exhibit any single-particle topological gap.

However, self-consistent Hartree-Fock calculations \cite{dong2024anomalous} reveal a qualitatively different ground state at an electron density $ n \approx 0.93 \times 10^{12}\,\mathrm{cm}^{-2}$, when the interlayer potential difference exceeds $\Delta < -25\,\mathrm{meV}$.
In this regime, the ground state is no longer a continuous translational-symmetric metallic state, but instead a charge-ordered state that spontaneously breaks graphene's continuous translational symmetry, and has a substantially lower energy than all translational-symmetric metallic solutions.

At filling $\nu=1$, the AHC exhibits a well-defined pattern of spontaneous symmetry breaking. 
First, time-reversal symmetry is spontaneously broken, with the system selecting a Chern number $C=\pm 1$, corresponding to the two possible orientations of the anomalous Hall response. 
Second, the conduction electrons form a periodic charge-ordered pattern in real space, reflecting the spontaneous breaking of continuous translational symmetry as shown in Figure \ref{fig:Theoretical}(d). Together, these two symmetry-breaking phenomena define the essential nature of the AHC.
Further calculations \cite{dong2024anomalous} show that the AHC is most stable relative to the metallic phase around $\Delta \approx -55\,\mathrm{meV}$ where the charge gap reaches its maximum and is consistent with experimentally estimated displacement fields. 
Notably, in the clean limit the AHC is a compressible state: the period of the interaction-driven electronic crystal is not locked by any external periodic potential. 
As a result, distortions corresponding to changes of the crystal period do not involve a finite pinning energy, reflecting the existence of a gapless sliding (phason) mode associated with spontaneous translational symmetry breaking. 
Nevertheless, even weak spatially varying potentials or disorder can pin the periodicity of the AHC \cite{tevsanovic1989hall}, lifting the degeneracy associated with continuous translations. 
Once pinned, the nontrivial Chern number can be revealed through the Středa formula and the quantized Hall conductance. 
Consequently, at filling $\nu=1$, both the topology and the charge gap are fundamentally determined by interaction effects, while the moiré potential does not generate the Chern-1 band but instead acts as a weak pinning field that stabilizes and reveals the AHC.

\subsection{The limitation of Hartree-Fock}
The limitations of the Hartree–Fock method in determining the ground state of the RMG/hBN system arise from multiple factors. Here, we discuss these factors point by point, as well as possible theoretical and experimental approaches for identifying the system’s ground state.

Firstly, the fundamental limitation lies in the restricted variational space of Hartree-Fock theory, which is confined to Slater determinant wave functions. 
Within this framework, the only metallic state accessible is the free-fermion Fermi liquid, while more general wave functions describing correlated metallic or non-Fermi-liquid phases are entirely excluded \cite{zhou2024fractional}.

This restricted variational space leads to a systematic bias of the Hartree-Fock method in favor of ordered states, particularly gapped insulating phases with broken symmetries. 
The self-consistent mean-field treatment can substantially lower the energy of symmetry-broken solutions, an effect that is typically absent for metallic states. 
As a result, within Hartree-Fock calculations, a translational-symmetry-breaking and gapped QAH crystal state generically wins over translational-invariant metallic solutions.

Secondly, the results of Hartree–Fock calculation sometimes depend strongly on the so-called ``reference field'' \cite{huang2024self, huang2025fractional}, which we did not mention in the above sections for the sake of simplicity. 
The reference field $ - V_{\text{HF}}(P_{\text{ref}})$ represents the interaction energy already present in the single-particle state before adding the explicit interaction term. Ref. \cite{huang2024self, huang2025fractional} systematically evaluated how different schemes of this reference field affect the convergence of the Hartree-Fock calculation and the resulting topological phases.

Lastly, even discarded the reference field, Ref. \cite{guo2024fractional} identified two competing candidate ground states at $\nu=1$, one topologically trivial ($C=0$) and the other topologically nontrivial ($C=1$), whose energies differ by only about $0.01 \sim 1,\mathrm{meV}$, as revealed by a renormalization-group analysis followed by Hartree–Fock calculations.

For these reason, in the weak or vanishing moiré potential limit, Hartree-Fock calculations cannot conclusively determine whether the AHC represents the true ground state of the system. 
While the ordered insulating state is found to have much lower energy than the free-fermion metallic solution, this does not rule out the possibility that a correlated metallic phase—absent from the Hartree-Fock variational space—may be energetically favorable in the real system. 
Consequently, the nature of the ground state in this regime remains an open question.

Based on these considerations, \cite{zhou2024fractional} pointed out that it is reasonable to expect that the QAH insulating \textit{ansatz} obtained from Hartree-Fock calculations represents the true ground state only when the external moiré superlattice potential exceeds a small critical threshold $V_M^c$. 
In this regime, the external superlattice pins the spontaneously formed crystalline order, removes the soft modes associated with continuous translational symmetry, and significantly enhances the stability of the ordered phase. Correspondingly, the reliability of the Hartree-Fock description is substantially improved.

To test this scenario experimentally, \cite{zhou2024fractional} proposed a controllable setup in which a moiré superlattice potential is induced in multilayer graphene via Coulomb interactions with a nearby twisted bilayer graphene (TBG), separated by a thin hBN spacer. 
In this scheme, the strength and spatial range of the induced potential can be continuously tuned by the hBN thickness and the carrier density in the TBG, while its lattice symmetry is determined by the TBG filling. 
Notably, a honeycomb moiré potential stabilizes the AHC, whereas a triangular superlattice suppresses it, underscoring the crucial role of the external superlattice symmetry in stabilizing the AHC. 
Ref. \cite{wang2025moire,ding2024engineering} fabricated similar devices consisting of bilayer graphene on superlattices, thereby demonstrating that the use of twisted bilayer structures to generate a tunable moiré potential is feasible.

Moreover, a different perspective on the moiré potentail has been addressed in Ref. \cite{guo2025correlationstabilizedanomaloushall}. By taking the realistic interaction within GW approximation, it has been found that the AHC does not likely happen in few-layer graphene system without the moiré potential. It highlights that the treatment of interaction beyond the Hartree-Fock is necessary to understand the physics here. 
Addtionally, Refs. \cite{dong2024stability, zeng2024sublattice, soejima2025lambda, song2024phase, song2024intertwined} studied the phase competition and transition among the AHC, the WCL, and the Fermi liquid states theoretically.

\subsection{Charge Density Waves}

On the moiré-distant side of the RMG/hBN moiré system, in addition to the Chern-1 isolated flat band appearing in an intermediate-coupling regime where interactions are significant while electrons remain quantum-mechanically extended (see Fig. \ref{fig:Theoretical}(d)), isolated bands with other Chern numbers are also observed. These features cannot be simply understood as direct consequences of single-particle moiré band structures.

Instead, a more natural interpretation is to place these states within a unified framework of interaction-driven charge density waves (CDW). 
From this perspective, the charge ordering and the associated topological properties originate primarily from electron-electron interactions rather than from single-particle effects of the moiré superlattice; 
the moiré potential does not generate the CDW itself, but mainly acts to pin, stabilize, and render visible interaction-dominated ground states that already exist in its absence.

\subsubsection{Chern-0 band and Wigner-like insulator}

As shown in Figure \ref{fig:Theoretical}(a), on the moiré-distant side ($\Delta < 0$), besides the Chern-1 anomalous Hall crystal (AHC) phase, there exists a wide region of insulating states at filling $\nu=1$ with vanishing Chern number, which is referred to as Wigner-like insulators (WLI) \cite{dong2024anomalous}.

The term “Wigner-like” emphasizes the similarity to the classical Wigner crystal \cite{wigner1934interaction}: electrons are strongly localized and form a regular real-space pattern, with a lattice constant primarily determined by the electron density. Individual electronic states are largely confined near their respective lattice sites, with a spatial extent much smaller than the average spacing between neighboring electrons.

WLI primarily appears at smaller twist angles compared to the AHC, corresponding to longer moiré lattice constants, lower electron densities at the same filling number, and smaller moiré Brillouin zones. 
These features substantially reduce the electronic kinetic energy and render electro-electron interactions relatively more dominant. 
As a result, it becomes energetically favorable for electrons to localize in real space and form an electronic crystal, maximizing inter-electron spacing to minimize Coulomb repulsion, rather than remaining in extended Bloch-like states. 
The resulting electronic crystal is also relatively rigid (note it is still compressible) against density fluctuations, naturally stabilizing a strongly localized Wigner-like insulating phase. 
Thus, it should be emphasized that this Wigner-like insulating phase is not a trivial band insulator, but an interaction-driven many-body insulating state characterized by a spontaneously formed electronic crystal structure.

This state can be viewed as an extreme form of a charge density wave (CDW), whose essential feature is the spontaneous breaking of continuous translational symmetry. 
In this regime, the low-energy excitations are dominated by collective displacement modes of the electronic crystal, such as phonons or the phase modes (phasons) of a CDW, while the electronic degrees of freedom associated with band topology are frozen out at low energies. 
Consequently, the low-energy effective theory in this parameter range no longer contains electronic degrees of freedom carrying nontrivial Chern numbers, and the insulating state is topologically trivial in the low-energy sector.

\subsubsection{Chern-2 band and topological CDW}

\begin{table}
\caption{Topological properties and symmetries of different electronic states.  }
\centering
\begin{tabular}{l c c c}
\hline
electronic state & Chern number & Time reversal symmetry  & Translational symmetry  \\
\hline
Wigner-like crystal              & zero      & preserved   &  Continuous symmetry  broken \\
Anomalous Hall crystal           & non-zero  & broken      & Continuous symmetry  broken \\
Charge density wave              & zero      & preserved   &  Discrete symmetry  broken \\
Topological charge density wave  & non-zero  & broken      & Discrete symmetry  broken  \\
\hline
\end{tabular}
\label{tab1}
\end{table}

In experimental works reported in Ref. \cite{ding2025electric}, band structures with higher Chern numbers, such as Chern-2, are also observed on the moiré-distant side at filling $\nu=1$ in R10G/hBN. To provide a unified understanding of these observations, it is necessary to re-examine the physical picture.

In previous studies, two simplifying assumptions were commonly adopted simultaneously: 
first, that the spontaneously formed electronic crystal has the same periodicity as the moiré lattice; 
and second, that only a single electron occupies each unit cell.
While these assumptions are physically intuitive, they are not essential. 
As discussed above, the electronic crystal is fundamentally generated by electron-electron interactions, and the corresponding electronic state is compressible in the clean limit. 
In this sense, the electronic crystal possesses an intrinsic tendency toward continuous translational symmetry breaking, with a lattice constant not rigidly fixed a priori.
The role of the moiré potential is therefore mainly to weakly pin the electronic crystal, rather than to rigidly fix either its lattice constant or the number of electrons per unit cell.

Consequently, as proposed in Ref. \cite{zhou2024fractional}, the basis vectors of the electronic crystal can be expressed as integer linear combinations of the moiré lattice vectors,
$\bm{R}_{\mathrm{crystal}} = m_1 \bm{R}_{\mathrm{Mo},1} + m_2 \bm{R}_{\mathrm{Mo},2}$ 
where $\bm{R}_{\mathrm{crystal}}$ and $\bm{R}_{\mathrm{Mo}}$ denote the basis vectors of the electronic crystal and moiré lattices, respectively, and $m_{1,2} \in \mathbb{N} $. 
For a triangular lattice, the corresponding lattice constant is given by 
$|\bm{R}_{\mathrm{crystal}}| = \sqrt{m_1^2 + m_2^2 + m_1 m_2} |\bm{R}_{\mathrm{Mo}}|$.
At a fixed electronic areal density $n$, one has $R^2 n \propto \nu$. 
Therefore, a filling factor $\nu$ defined with respect to the moiré lattice is equivalent to an effective filling
$\nu^\prime = \nu (m_1^2 + m_2^2 + m_1 m_2)$
in an electronic crystal with lattice constant $R_{\mathrm{crystal}}$. 
In other words, multiple electrons can reside within a single electronic unit cell, collectively forming a new interaction-driven many-body quantum state, which has been referred to as a QAHC-$\nu'$ state in \cite{zhou2025new}.

This perspective further implies that, even at the same electronic density, multiple quantum states characterized by different electronic lattice constants and different numbers of electrons per unit cell may energetically compete. Such competition provides a natural explanation for the emergence of multiple anomalous Hall phases in the moiré-distant regime.

\section{Discussion}
\label{sec:discussion}

The studies reviewed in this paper collectively establish the RMG/hBN moiré system as a versatile platform for realizing and manipulating a broad spectrum of quantum phases. We have mainly focused on the QAH emergent at the integer filling $\nu=1$. The role and interplay of moiré potential and electron-electron interaction have been addressed in detail.

Due to the length of the paper, there are several topics that are not included. For example,
the FQAH states observed at fractional filling ($\nu = 2/3, 3/5, \dots$) of a Chern-1 band. 
Thus, a task for theory is to first explain the emergence of the FQAH phases in the RMG/hBN system.
At partial rational fillings of the valley polarized bands, theoretical calculations using exact diagonalization have shown
that topological phases equivalent to FQH states develop \cite{dong2024theory}. The theoretical backgrounds shown in Sec. 3 provide a microscopic basis for the observation of FQAH physics in the RMG/hBN systems.
Another example is,
besides the RMG/hBN systems, very recently a new ($M+N$) L moiré system, twisted rhombohedral $M$- and $N$-layer graphene, has been successfully synthesized \cite{Wang2026twistRG3+3, Dong2025twistRG2+3, Liu2025twistRG1+5, Li2025twistRG2+4, Wang2026twistRG1+345, chen2026layerengineeredquantumanomaloushall}. 
In addition, RMG on monolayer TMD substrates has also been realized \cite{han2024large, sha2024observation}, where the spin–orbit coupling induced by the TMD strongly modifies the electronic bands even in the absence of a moiré superlattice.
They are highly tunable platforms for achieving and controlling high-Chern-number QAH insulators, with the demonstration of the topological Chern number can be precisely engineered and dynamically switched through layer configuration, electrostatic doping, and applied displacement fields \cite{Liu2025twistRG1+5, Wang2026twistRG1+345, chen2026layerengineeredquantumanomaloushall, liu2025layer}.

Moreover, despite rapid progress, several open issues remain \cite{uchida2025non, liu2025symmetry, kudo2024quantum, gong2025instabilities, tan2024parent}. In R4G/hBN and R5G/hBN moiré superlattice devices, anomalous Hall hysteresis has been observed near fillings around \(\nu \approx -2.5\) \cite{lu2025extended,choi2025superconductivity,Matthew2025R5GPRX}, yet the quantized values inferred from transport are sometimes inconsistent with Chern numbers extracted from the Streda analysis \cite{streda1982theory,lu2025extended,Matthew2025R5GPRX}. The microscopic origin of this discrepancy remains unresolved.

Another open issue is the recently reported EQAH phenomenology, where Hall quantization persists over a continuous carrier-density window instead of a single commensurate filling \cite{lu2025extended, xiang2025continuously, JCCMP2024Jan}. This phenomenology goes beyond the simplest picture of an isolated Chern gap at one commensurate filling. 
The microscopic mechanism behind such a continuous QAH window has been actively discussed \cite{zhou2024fractional,dong2024anomalous,dong2024theory,dong2024stability,patri2024extended}.
Two possible mechanisms have been proposed for the extended quantization. It may reflect spontaneous translational-symmetry breaking at the moir\'e scale, forming a charge-ordered AHC that preserves quantized $R_{xy}$ across a range of fillings \cite{zhou2024fractional,dong2024anomalous,dong2024stability}.
Alternatively, it may correspond to an IQAH background at a nearby commensurate filling, with excess holes forming a pinned Wigner/CDW crystal, reminiscent of re-entrant QH phenomenology \cite{patri2024extended}.

Looking forward, key challenges include enhancing the robustness of FQAH states (e.g.\ by engineering larger excitation gaps, potentially through proximity-induced spin--orbit coupling). It is also important to develop gate-defined junctions that enable controlled coupling between QAH phases and superconductivity in RMG devices, enabling systematic tests of their interfacial physics \cite{lu2024fractional,lu2025extended,choi2025superconductivity,kumar2025superconductivity,Matthew2025RGSC,ARCMP2025review}. 
More broadly, the high tunability of graphene provides a practical route to explore correlated and topological phases in RMG by systematically varying the layer number and the twist angle, which directly reshape the bandwidth, moir\'e length scale, and the balance between competing orders \cite{lu2024fractional,ding2025electric,Xiaobo2025orientation}. 
The generic model for AHC in the moiré superlattice systems has yet to be discovered \cite{xie2025generic, lu2025generic}. 
Researchers are also employing state-of-the-art tools, such as artificial intelligence, to investigate the relevant problems \cite{valenti2025quantum, teng2025solving}.

\ \\
\ \\

\clearpage 
\section{Appendix}

\subsection{derive Hartree-Fock approximation}

Rewrite Eq.\,\ref{eq:intra},
\begin{equation}\label{eq:re_intra}
\begin{aligned}
   \hat{H}_{\text{intra}} 
   &= \frac{1}{2\mathcal{V}} \sum_{\tau_1, \tau_2, s_1 ,s_2}  
   \sum_{\{ \bm{k}_i\} } \sum_{ \{n_i \} }
   V^{\tau_2 \tau_1}_{n_4 n_3 n_2 n_1}(\bm{k}_1,\bm{k}_2,\bm{k}_3, \bm{k}_4)
     \hat{\psi}^{\dagger}_{\bm{k}_4 s_1 \tau_1  n_4}
     \hat{\psi}^{\dagger}_{\bm{k}_3 s_2 \tau_2  n_3}
     \hat{\psi}_{\bm{k}_2 s_2 \tau_2  n_2}  
     \hat{\psi}_{\bm{k}_1 s_1 \tau_1  n_1}  \\
\end{aligned}
\end{equation}
where
\begin{equation}
\begin{aligned}
   V^{\tau_2 \tau_1}_{n_4 n_3 n_2 n_1}(\bm{k}_1,\bm{k}_2,\bm{k}_3, \bm{k}_4) &= \sum_{\bm{Q}} V(\bm{q}+\bm{Q})  \Lambda^{\tau_2, +}_{\bm{q}+\bm{Q}, \bm{k}_3, \bm{k}_2; n_3, n_2}
  \left( \Lambda^{\tau_1, +}_{\bm{q}+\bm{Q}, \bm{k}_1, \bm{k}_4; n_1, n_4} \right)^{*}
\end{aligned}
\end{equation}
Apply indices substitution: $\bm{k}_1 \leftrightarrow \bm{k}_2$, $\bm{k}_3 \leftrightarrow \bm{k}_4$, ${\tilde{\bm{q}}}\leftrightarrow -{\tilde{\bm{q}}}$, $\tau_1 \leftrightarrow \tau_2$, $n_1 \leftrightarrow n_2$, $n_3 \leftrightarrow n_4$ then
\begin{equation}
\begin{aligned}
   V^{\tau_1 \tau_2}_{n_3 n_4 n_1 n_2}(\bm{k}_2,\bm{k}_1,\bm{k}_4, \bm{k}_3) 
   &= \sum_{-\bm{Q}} V(-\bm{q}-\bm{Q})  \Lambda^{\tau_2, +}_{-\bm{q}-\bm{Q}, \bm{k}_4, \bm{k}_1; n_4, n_1}
  \left( \Lambda^{\tau_1, +}_{-\bm{q}-\bm{Q}, \bm{k}_2, \bm{k}_3; n_2, n_3} \right)^{*}  \\
   &= \sum_{\bm{Q}} V(\bm{q}+\bm{Q}) \left( \Lambda^{\tau_2, +}_{\bm{q}+\bm{Q}, \bm{k}_1, \bm{k}_4; n_1, n_4}  \right)^{*}
   \Lambda^{\tau_1, +}_{\bm{q}+\bm{Q}, \bm{k}_3, \bm{k}_2; n_3, n_2}  \\
   &= V^{\tau_2 \tau_1}_{n_4 n_3 n_2 n_1}(\bm{k}_1,\bm{k}_2,\bm{k}_3, \bm{k}_4) 
\end{aligned}
\end{equation}
In addition, we define
\begin{equation}\label{}
\begin{aligned}
   V^{\tau_2 \tau_1, \text{H}}_{n_4 n_3 n_2 n_1}(\bm{k}_1,\bm{k}_2) &:= V^{\tau_2 \tau_1}_{n_4 n_3 n_2 n_1}(\bm{k}_1,\bm{k}_2,\bm{k}_2, \bm{k}_1) \\
   V^{\tau_2 \tau_1, \text{F}}_{n_4 n_3 n_2 n_1}(\bm{k}_1,\bm{k}_2) &:= V^{\tau_2 \tau_1}_{n_4 n_3 n_2 n_1}(\bm{k}_1,\bm{k}_2,\bm{k}_1, \bm{k}_2)
\end{aligned}
\end{equation}
for further use. Temporarily omitting some subscripts, we have
\begin{equation}
\begin{aligned}
  & \hat{\psi}^\dagger_{\bm{k}_4} \hat{\psi}^\dagger_{\bm{k}_3} \hat{\psi}_{\bm{k}_2} \hat{\psi}_{\bm{k}_1} \\
  =&  \delta(\hat{\psi}^\dagger_{\bm{k}_4} \hat{\psi}^\dagger_{\bm{k}_3} \hat{\psi}_{\bm{k}_2} \hat{\psi}_{\bm{k}_1}) 
  + \langle \hat{\psi}^\dagger_{\bm{k}_4} \hat{\psi}_{\bm{k}_1} \rangle \delta(\hat{\psi}^\dagger_{\bm{k}_3} \hat{\psi}_{\bm{k}_2}) + \langle \hat{\psi}^\dagger_{\bm{k}_3} \hat{\psi}_{\bm{k}_2} \rangle \delta(\hat{\psi}^\dagger_{\bm{k}_4} \hat{\psi}_{\bm{k}_1})  
  -  \langle \hat{\psi}^\dagger_{\bm{k}_4} \hat{\psi}_{\bm{k}_2} \rangle \delta(\hat{\psi}^\dagger_{\bm{k}_3} \hat{\psi}_{\bm{k}_1})  \\
  &- \langle \hat{\psi}^\dagger_{\bm{k}_3} \hat{\psi}_{\bm{k}_1} \rangle \delta(\hat{\psi}^\dagger_{\bm{k}_4} \hat{\psi}_{\bm{k}_2}) 
  +  \langle \hat{\psi}^\dagger_{\bm{k}_4} \hat{\psi}_{\bm{k}_1} \rangle \langle \hat{\psi}^\dagger_{\bm{k}_3} \hat{\psi}_{\bm{k}_2} \rangle - \langle \hat{\psi}^\dagger_{\bm{k}_4} \hat{\psi}_{\bm{k}_2} \rangle \langle \hat{\psi}^\dagger_{\bm{k}_3} \hat{\psi}_{\bm{k}_1} \rangle \\
  \approx &  \langle \hat{\psi}^\dagger_{\bm{k}_4} \hat{\psi}_{\bm{k}_1} \rangle \hat{\psi}^\dagger_{\bm{k}_3} \hat{\psi}_{\bm{k}_2} + \langle \hat{\psi}^\dagger_{\bm{k}_3} \hat{\psi}_{\bm{k}_2} \rangle \hat{\psi}^\dagger_{\bm{k}_4} \hat{\psi}_{\bm{k}_1} 
  -  \langle \hat{\psi}^\dagger_{\bm{k}_4} \hat{\psi}_{\bm{k}_2} \rangle \hat{\psi}^\dagger_{\bm{k}_3} \hat{\psi}_{\bm{k}_1} \\
  & - \langle \hat{\psi}^\dagger_{\bm{k}_3} \hat{\psi}_{\bm{k}_1} \rangle \hat{\psi}^\dagger_{\bm{k}_4} \hat{\psi}_{\bm{k}_2}   
  - \langle \hat{\psi}^\dagger_{\bm{k}_4} \hat{\psi}_{\bm{k}_1} \rangle \langle \hat{\psi}^\dagger_{\bm{k}_3} \hat{\psi}_{\bm{k}_2} \rangle +   \langle \hat{\psi}^\dagger_{\bm{k}_4} \hat{\psi}_{\bm{k}_2} \rangle \langle \hat{\psi}^\dagger_{\bm{k}_3} \hat{\psi}_{\bm{k}_1} \rangle  \\
  = &  P(\bm{k}_4, \bm{k}_1) \hat{\psi}^\dagger_{\bm{k}_3} \hat{\psi}_{\bm{k}_2} 
  + P(\bm{k}_3, \bm{k}_2) \hat{\psi}^\dagger_{\bm{k}_4} \hat{\psi}_{\bm{k}_1} 
 -  P(\bm{k}_4, \bm{k}_2) \hat{\psi}^\dagger_{\bm{k}_3} \hat{\psi}_{\bm{k}_1} \\
 & - P(\bm{k}_3, \bm{k}_1) \hat{\psi}^\dagger_{\bm{k}_4} \hat{\psi}_{\bm{k}_2}   
 -  P(\bm{k}_4, \bm{k}_1) P(\bm{k}_3, \bm{k}_2) 
 +   P(\bm{k}_4, \bm{k}_2) P(\bm{k}_3, \bm{k}_1) \\
  = & \delta_{\bm{k}_1,\bm{k}_4} P(\bm{k}_1) \hat{\psi}^\dagger_{\bm{k}_3} \hat{\psi}_{\bm{k}_2} 
  + \delta_{\bm{k}_2,\bm{k}_3} P(\bm{k}_2) \hat{\psi}^\dagger_{\bm{k}_4} \hat{\psi}_{\bm{k}_1} 
 -  \delta_{\bm{k}_2,\bm{k}_4} P(\bm{k}_2) \hat{\psi}^\dagger_{\bm{k}_3} \hat{\psi}_{\bm{k}_1} \\
   & - \delta_{\bm{k}_1,\bm{k}_3} P(\bm{k}_1) \hat{\psi}^\dagger_{\bm{k}_4} \hat{\psi}_{\bm{k}_2}   
 -  \delta_{\bm{k}_1,\bm{k}_4} P(\bm{k}_1) \delta_{\bm{k}_2,\bm{k}_3} P(\bm{k}_2) 
 +   \delta_{\bm{k}_2,\bm{k}_4} P(\bm{k}_2) \delta_{\bm{k}_1,\bm{k}_3} P(\bm{k}_1) \\
\end{aligned}
\end{equation}
where the first equality uses the Wick theorem, and the approximate equality substitutes the definition $\delta\hat{O} := \hat{O} - \langle \hat{O} \rangle$ and omits higher-order terms. Thus, Eq.\,\ref{eq:re_intra} can be decoupled into three parts: Hartree term $\hat{H}_{\text{intra}}^{\text{H}}$, Fock term $\hat{H}_{\text{intra}}^{\text{F}}$ and constant (condensed energy) term $-E_{\text{intra}}^{\text{C}}$.

The Hartree term ($\delta_{\bm{k}_1,\bm{k}_4} = \delta_{\bm{k}_2,\bm{k}_3} = \delta_{\bm{q}, 0}$):
\begin{equation}
\begin{aligned}
\hat{H}_{\text{intra}}^{\text{H}}
&= \frac{1}{2\mathcal{V}} \sum_{\tau_1, \tau_2, s_1 ,s_2}
\sum_{\{ \bm{k}_i\} } \sum_{ \{n_i \} }
V^{\tau_2 \tau_1}_{n_4 n_3 n_2 n_1}(\bm{k}_1,\bm{k}_2,\bm{k}_3, \bm{k}_4)
\times \\
& \quad \left[ \delta_{\bm{k}_1,\bm{k}_4} P(\bm{k}_1)_{s_1 \tau_1 n_4, s_1 \tau_1 n_1 } \hat{\psi}^\dagger_{\bm{k}_3 s_2 \tau_2  n_3} \hat{\psi}_{\bm{k}_2 s_2 \tau_2 n_2}
+ \delta_{\bm{k}_2,\bm{k}_3} P(\bm{k}_2)_{s_2 \tau_2 n_3, s_2 \tau_2 n_2 } \hat{\psi}^\dagger_{\bm{k}_4 s_1 \tau_1  n_4} \hat{\psi}_{\bm{k}_1 s_1 \tau_1  n_1}  \right] \\
&= \frac{1}{2\mathcal{V}} \sum_{\tau_1, \tau_2, s_1 ,s_2}
\sum_{\bm{k}_1, \bm{k}_2 } \sum_{ \{n_i \} }
\left[ V^{\tau_2 \tau_1, \text{H}}_{n_4 n_3 n_2 n_1}(\bm{k}_1,\bm{k}_2) P(\bm{k}_1)_{s_1 \tau_1 n_4, s_1 \tau_1 n_1} \hat{\psi}^\dagger_{\bm{k}_2 s_2 \tau_2 n_3} \hat{\psi}_{\bm{k}_2 s_2 \tau_2 n_2} \right. \\
& \left. \text{\qquad\qquad\qquad\qquad\qquad\qquad}
+ V^{\tau_2 \tau_1, \text{H}}_{n_4 n_3 n_2 n_1}(\bm{k}_1,\bm{k}_2) P(\bm{k}_2)_{s_2 \tau_2 n_3, s_2 \tau_2 n_2} \hat{\psi}^\dagger_{\bm{k}_1 s_1 \tau_1 n_4} \hat{\psi}_{\bm{k}_1 s_1 \tau_1 n_1} \right]  \\
&= \frac{1}{2\mathcal{V}} \sum_{\tau_1, \tau_2, s_1 ,s_2}
\sum_{\bm{k}_1, \bm{k}_2 } \sum_{ \{n_i \} }
\left[ V^{\tau_1 \tau_2, \text{H}}_{n_3 n_4 n_1 n_2}(\bm{k}_2,\bm{k}_1) P(\bm{k}_1)_{s_1 \tau_1 n_4, s_1 \tau_1 n_1} \hat{\psi}^\dagger_{\bm{k}_2 s_2 \tau_2 n_3} \hat{\psi}_{\bm{k}_2 s_2 \tau_2 n_2} \right. \\
& \left. \text{\qquad\qquad\qquad\qquad\qquad\qquad}
+ V^{\tau_2 \tau_1, \text{H}}_{n_4 n_3 n_2 n_1}(\bm{k}_1,\bm{k}_2) P(\bm{k}_2)_{s_2 \tau_2 n_3, s_2 \tau_2 n_2} \hat{\psi}^\dagger_{\bm{k}_1 s_1 \tau_1 n_4} \hat{\psi}_{\bm{k}_1 s_1 \tau_1 n_1} \right]  \\
&= \frac{1}{2\mathcal{V}} \sum_{\tau_1, \tau_2, s_1 ,s_2}
\sum_{\bm{k}_1, \bm{k}_2 } \sum_{ \{n_i \} }
\left[ \underbrace{V^{\tau_2 \tau_1, \text{H}}_{n_1 n_3 n_4 n_2}(\bm{k}_1,\bm{k}_2) P(\bm{k}_2)_{s_2 \tau_2 n_3, s_2 \tau_2 n_4} \hat{\psi}^\dagger_{\bm{k}_1 s_1 \tau_1 n_1} \hat{\psi}_{\bm{k}_1 s_1 \tau_1 n_2}}_{\text{by substitution: }\bm{k}_1 \leftrightarrow \bm{k}_2, \tau_1 \leftrightarrow \tau_2, s_1 \leftrightarrow s_2,  n_1 \to n_4 \to n_3 \to n_1} \right. \\
& \left. \text{\qquad\qquad\qquad\qquad\qquad\qquad}
+ \underbrace{V^{\tau_2 \tau_1, \text{H}}_{n_1 n_3 n_4 n_2}(\bm{k}_1,\bm{k}_2) P(\bm{k}_2)_{s_2 \tau_2 n_3, s_2 \tau_2 n_4} \hat{\psi}^\dagger_{\bm{k}_1 s_1 \tau_1 n_1} \hat{\psi}_{\bm{k}_1 s_1 \tau_1 n_2} }_{\text{by substitution: } n_1 \to n_2 \to n_4 \to n_1 } \right]  \\
&= \frac{1}{\mathcal{V}} \sum_{\tau_1, \tau_2, s_1 ,s_2}
\sum_{\bm{k}_1, \bm{k}_2 } \sum_{ \{n_i \} }
V^{\tau_2 \tau_1, \text{H}}_{n_1 n_3 n_4 n_2}(\bm{k}_1,\bm{k}_2) P(\bm{k}_2)_{s_2 \tau_2 n_3, s_2 \tau_2 n_4} \hat{\psi}^\dagger_{\bm{k}_1 s_1 \tau_1 n_1} \hat{\psi}_{\bm{k}_1 s_1 \tau_1 n_2}  \\
\end{aligned}
\end{equation}

The Fock term ($\delta_{\bm{k}_1,\bm{k}_3} = \delta_{\bm{k}_2,\bm{k}_4}$):
\begin{equation}
\begin{aligned}
\hat{H}_{\text{intra}}^{\text{F}}
&= \frac{1}{2\mathcal{V}} \sum_{\tau_1, \tau_2, s_1 ,s_2}
\sum_{\{ \bm{k}_i\} } \sum_{ \{n_i \} }
V^{\tau_2 \tau_1}_{n_4 n_3 n_2 n_1}(\bm{k}_1,\bm{k}_2,\bm{k}_3, \bm{k}_4)
\times \\
& \quad \left[ -  \delta_{\bm{k}_2,\bm{k}_4} P(\bm{k}_2)_{s_1 \tau_1  n_4, s_2 \tau_2 n_2} \hat{\psi}^\dagger_{\bm{k}_3 s_2 \tau_2  n_3} \hat{\psi}_{\bm{k}_1 s_1 \tau_1  n_1} 
   - \delta_{\bm{k}_1,\bm{k}_3} P(\bm{k}_1)_{s_2 \tau_2  n_3, s_1 \tau_1  n_1} \hat{\psi}^\dagger_{\bm{k}_4 s_1 \tau_1  n_4} \hat{\psi}_{\bm{k}_2 s_2 \tau_2 n_2}   \right] \\
&= - \frac{1}{2\mathcal{V}} \sum_{\tau_1, \tau_2, s_1 ,s_2}
\sum_{\bm{k}_1, \bm{k}_2 } \sum_{ \{n_i \} }
\left[ V^{\tau_2 \tau_1, \text{F}}_{n_4 n_3 n_2 n_1}(\bm{k}_1,\bm{k}_2) P(\bm{k}_2)_{s_1 \tau_1  n_4, s_2 \tau_2 n_2} \hat{\psi}^\dagger_{\bm{k}_1 s_2 \tau_2  n_3} \hat{\psi}_{\bm{k}_1 s_1 \tau_1  n_1}  \right. \\
& \left. \text{\qquad\qquad\qquad\qquad\qquad\qquad}
+ V^{\tau_2 \tau_1, \text{F}}_{n_4 n_3 n_2 n_1}(\bm{k}_1,\bm{k}_2) P(\bm{k}_1)_{s_2 \tau_2  n_3, s_1 \tau_1  n_1} \hat{\psi}^\dagger_{\bm{k}_2 s_1 \tau_1  n_4} \hat{\psi}_{\bm{k}_2 s_2 \tau_2 n_2}  \right]  \\
&= - \frac{1}{2\mathcal{V}} \sum_{\tau_1, \tau_2, s_1 ,s_2}
\sum_{\bm{k}_1, \bm{k}_2 } \sum_{ \{n_i \} }
\left[ V^{\tau_2 \tau_1, \text{F}}_{n_4 n_3 n_2 n_1}(\bm{k}_1,\bm{k}_2) P(\bm{k}_2)_{s_1 \tau_1  n_4, s_2 \tau_2 n_2} \hat{\psi}^\dagger_{\bm{k}_1 s_2 \tau_2  n_3} \hat{\psi}_{\bm{k}_1 s_1 \tau_1  n_1}  \right. \\
& \left. \text{\qquad\qquad\qquad\qquad\qquad\qquad}
+ V^{\tau_1 \tau_2, \text{F}}_{n_3 n_4 n_1 n_2}(\bm{k}_2,\bm{k}_1) P(\bm{k}_1)_{s_2 \tau_2  n_3, s_1 \tau_1  n_1} \hat{\psi}^\dagger_{\bm{k}_2 s_1 \tau_1  n_4} \hat{\psi}_{\bm{k}_2 s_2 \tau_2 n_2}  \right]  \\
&= - \frac{1}{2\mathcal{V}} \sum_{\tau_1, \tau_2, s_1 ,s_2}
\sum_{\bm{k}_1, \bm{k}_2 } \sum_{ \{n_i \} }
\left[ \underbrace{V^{\tau_1 \tau_2, \text{F}}_{n_3 n_1 n_4 n_2}(\bm{k}_1,\bm{k}_2) P(\bm{k}_2)_{s_2 \tau_2  n_3, s_1 \tau_1 n_4} \hat{\psi}^\dagger_{\bm{k}_1 s_1 \tau_1  n_1} \hat{\psi}_{\bm{k}_1 s_2 \tau_2  n_2}}_{\text{by substitution: } \tau_1 \leftrightarrow \tau_2, s_1 \leftrightarrow s_2,  n_1\to n_2 \to n_4 \to n_3 \to n_1 } \right. \\
& \left. \text{\qquad\qquad\qquad\qquad\qquad\qquad}
+ \underbrace{V^{\tau_1 \tau_2, \text{F}}_{n_3 n_1 n_4 n_2}(\bm{k}_1,\bm{k}_2) P(\bm{k}_2)_{s_2 \tau_2  n_3, s_1 \tau_1  n_4} \hat{\psi}^\dagger_{\bm{k}_1 s_1 \tau_1  n_1} \hat{\psi}_{\bm{k}_1 s_2 \tau_2 n_2}  }_{\text{by substitution: } \bm{k}_1 \leftrightarrow \bm{k}_2,  n_4 \leftrightarrow n_1} \right]  \\
&= - \frac{1}{\mathcal{V}} \sum_{\tau_1, \tau_2, s_1 ,s_2}
\sum_{\bm{k}_1, \bm{k}_2 } \sum_{ \{n_i \} }
V^{\tau_1 \tau_2, \text{F}}_{n_3 n_1 n_4 n_2}(\bm{k}_1,\bm{k}_2) P(\bm{k}_2)_{s_2 \tau_2  n_3, s_1 \tau_1 n_4} \hat{\psi}^\dagger_{\bm{k}_1 s_1 \tau_1  n_1} \hat{\psi}_{\bm{k}_1 s_2 \tau_2  n_2}   \\
\end{aligned}
\end{equation}

The condensed energy:
\begin{equation}
\begin{aligned}
E_{\text{intra}}^{\text{C}}
&= \frac{1}{2\mathcal{V}} \sum_{\tau_1, \tau_2, s_1 ,s_2}
\sum_{\{ \bm{k}_i\} } \sum_{ \{n_i \} }
V^{\tau_2 \tau_1}_{n_4 n_3 n_2 n_1}(\bm{k}_1,\bm{k}_2,\bm{k}_3, \bm{k}_4)
\times  \\
& \text{\qquad\qquad\qquad\qquad\qquad} \left[ \delta_{\bm{k}_1,\bm{k}_4} P(\bm{k}_1)_{s_1 \tau_1 n_4, s_1 \tau_1 n_1 } \delta_{\bm{k}_2,\bm{k}_3} P(\bm{k}_2)_{s_2 \tau_2 n_3, s_2 \tau_2 n_2 }
 \right. \\
&  \text{\qquad\qquad\qquad\qquad\qquad\quad} \left.  - \delta_{\bm{k}_2,\bm{k}_4} P(\bm{k}_2)_{s_1 \tau_1  n_4, s_2 \tau_2 n_2} \delta_{\bm{k}_1,\bm{k}_3} P(\bm{k}_1)_{s_2 \tau_2  n_3, s_1 \tau_1  n_1} \right] \\
&= \frac{1}{2\mathcal{V}} \sum_{\tau_1, \tau_2, s_1 ,s_2}
\sum_{\{ \bm{k}_i\} } \sum_{ \{n_i \} }
\left[ V^{\tau_2 \tau_1, \text{H}}_{n_4 n_3 n_2 n_1}(\bm{k}_1,\bm{k}_2) P(\bm{k}_1)_{s_1 \tau_1 n_4, s_1 \tau_1 n_1 } P(\bm{k}_2)_{s_2 \tau_2 n_3, s_2 \tau_2 n_2 }
 \right. \\
&  \text{\qquad\qquad\qquad\qquad\qquad\quad} \left.  - V^{\tau_2 \tau_1, \text{F}}_{n_4 n_3 n_2 n_1}(\bm{k}_1,\bm{k}_2) P(\bm{k}_2)_{s_1 \tau_1  n_4, s_2 \tau_2 n_2}  P(\bm{k}_1)_{s_2 \tau_2  n_3, s_1 \tau_1  n_1} \right] \\
\end{aligned}
\end{equation}

Besides, in numerical calculations, Hartree and Fock terms can be combined as
\begin{equation}
\begin{aligned}
& \hat{H}_{\text{intra}}^{\text{H}} + \hat{H}_{\text{intra}}^{\text{F}}  \\ 
=& \frac{1}{\mathcal{V}} \sum_{\tau_1,  s_1 }
\sum_{\bm{k}_1, \bm{k}_2 } \sum_{ \{n_i \} }
\left[ \sum_{\tau_3, s_3} V^{\tau_3 \tau_1, \text{H}}_{n_1 n_3 n_4 n_2}(\bm{k}_1,\bm{k}_2) P(\bm{k}_2)_{s_3 \tau_3 n_3, s_3 \tau_3 n_4} \right] \sum_{\tau_2, s_2} \delta_{s_1 , s_2} \delta_{\tau_1, \tau_2}\hat{\psi}^\dagger_{\bm{k}_1 s_1 \tau_1 n_1} \hat{\psi}_{\bm{k}_1 s_2 \tau_2 n_2}  \\
 & - \frac{1}{\mathcal{V}} \sum_{\tau_1, \tau_2, s_1 ,s_2}
\sum_{\bm{k}_1, \bm{k}_2 } \sum_{ \{n_i \} }
V^{\tau_1 \tau_2, \text{F}}_{n_3 n_1 n_4 n_2}(\bm{k}_1,\bm{k}_2) P(\bm{k}_2)_{s_2 \tau_2  n_3, s_1 \tau_1 n_4} \hat{\psi}^\dagger_{\bm{k}_1 s_1 \tau_1  n_1} \hat{\psi}_{\bm{k}_1 s_2 \tau_2  n_2}   \\
  =& \frac{1}{\mathcal{V}} \sum_{\bm{k}_1} \sum_{\tau_1, \tau_2, s_1 ,s_2}
   \sum_{ n_1, n_2 }
  \hat{\psi}^\dagger_{\bm{k}_1 s_1 \tau_1 n_1} 
  h_{\text{HF}}(\bm{k}_1)_{ s_1 \tau_1 n_1,  s_2 \tau_2 n_2}
  \hat{\psi}_{\bm{k}_1 s_2 \tau_2 n_2} 
\end{aligned}
\end{equation}
where
\begin{equation}
\begin{aligned}
  h_{\text{HF}}(\bm{k}_1)_{ s_1 \tau_1 n_1,  s_2 \tau_2 n_2} &= \sum_{ \bm{k}_2 }  \sum_{ n_3, n_4 } \left[ \delta_{s_1 , s_2} \delta_{\tau_1, \tau_2} \sum_{\tau_3, s_3} V^{\tau_3 \tau_1, \text{H}}_{n_1 n_3 n_4 n_2}(\bm{k}_1,\bm{k}_2) P(\bm{k}_2)_{s_3 \tau_3 n_3, s_3 \tau_3 n_4} \right.  \\
  & \text{\qquad\qquad\qquad}  \left.
   - V^{\tau_1 \tau_2, \text{F}}_{n_3 n_1 n_4 n_2}(\bm{k}_1,\bm{k}_2) P(\bm{k}_2)_{s_2 \tau_2  n_3, s_1 \tau_1 n_4}   \right] \\
\end{aligned}
\end{equation}

\subsection{Numerical Calculation of Chern Number}

The quantum geometric tensor, a Hermite tensor, is defined as
\begin{equation}
Q_{\mu\nu}(\bm{k})
= \langle \partial_\mu \psi_{\bm{k}} | \left( I - |\psi_{\bm{k}} \rangle \langle \psi_{\bm{k}} | \right) | \partial_\nu \psi_{\bm{k}} \rangle,
\end{equation}
from which the Berry curvature $F_{\mu\nu}(\bm{k})$ and the quantum metric $g_{\mu\nu}(\bm{k})$ are extracted as
\begin{equation}
\begin{aligned}
F_{\mu\nu}(\bm{k}) &= -2 \, \mathrm{Im} \, Q_{\mu\nu}(\bm{k}), &
g_{\mu\nu}(\bm{k}) &= \mathrm{Re} \, Q_{\mu\nu}(\bm{k}).
\end{aligned}
\end{equation}
Here, $\partial_\mu =\frac{\partial}{\partial k_\mu}$, and $\mu$, $\nu$ denote arbitrary directions in momentum space, which need not be orthogonal.

Equivalently, the Berry curvature can be expressed in terms of the Berry connection $\mathcal{A}_{\mu}(\bm{k}) = i \langle\psi_{\bm{k}} | \partial_\mu \psi_{\bm{k}} \rangle$ as
\begin{equation}
F_{\mu\nu}(\bm{k}) = \partial_{\mu} \mathcal{A}_{\nu}(\bm{k}) - \partial_{\nu} \mathcal{A}_{\mu}(\bm{k}).
\end{equation}
In three dimensions, this relation takes the form $\bm{F} = \nabla \times \bm{\mathcal{A}}$, analogous to the relation between a magnetic field and its vector potential. Note that $\bm{F}$ is gauge-invariant, while $\bm{\mathcal{A}}$ is not. The Stokes theorem connects the surface integral of $\bm{F}$ to the line integral of $\bm{\mathcal{A}}$ around the boundary.

The Chern number $C$ of an isolated band is given by the integral of the Berry curvature over the moir\'e Brillouin zone ($\mathrm{mBZ}$):
\begin{equation}
C = \frac{1}{2\pi} \int_{\mathrm{mBZ}} F_{xy}(\bm{k}) \, \mathrm{d}^2\bm{k} = \frac{1}{2\pi} \int_{\mathrm{mBZ}} \bm{F}(\bm{k}) \cdot \mathrm{d}\bm{S}.
\end{equation}

\subsubsection{Discretization and Wilson Loop Method}

To compute the Chern number numerically, the (moir\'e) Brillouin zone is discretized into a mesh of small plaquettes $\{P_i\}$. The total Berry flux is approximated by summing the fluxes through each plaquette:
\begin{equation}
2\pi C = \int_{\mathrm{mBZ}} \bm{F}(\bm{k}) \cdot \mathrm{d}\bm{S} \approx \sum_{i=1}^{N_p} \Phi_i,
\end{equation}
where $\Phi_i$ is the Berry flux through the $i$-th plaquette $P_i$. Using Stokes' theorem, the flux through each plaquette is given by the Berry phase accumulated along its boundary,
\begin{equation}
\Phi_{i} = \iint_{P_i} \bm{F}(\bm{k}) \cdot \mathrm{d}\bm{S} = \iint_{P_i} (\nabla \times \bm{\mathcal{A}}) \cdot \mathrm{d}\bm{S} = \oint_{\partial P_i} \bm{\mathcal{A}} \cdot \mathrm{d}\bm{k}.
\end{equation}

The line integral along the boundary $\partial P_i$ is evaluated using the Wilson loop method. For a small displacement from $\bm{k}_i$ to $\bm{k}_j$, the link variable is defined as
\begin{equation}
U_{i,j} = \langle \psi_{\bm{k}_i} | \psi_{\bm{k}_j} \rangle.
\end{equation}
Expanding $|\psi_{\bm{k}_j}\rangle$ to first order in $\delta \bm{k} = \bm{k}_j - \bm{k}_i$ gives
\begin{equation}
\begin{aligned}
U_{i,j} &= \langle \psi_{\bm{k}_i} | \left( |\psi_{\bm{k}_i} \rangle + \delta k_{\mu} |\partial_{\mu}\psi_{\bm{k}_i} \rangle + \mathcal{O} ((\delta k_{\mu})^2) \right) \\
&= 1 + \delta k_{\mu} \langle \psi_{\bm{k}_i} | \partial_{\mu}\psi_{\bm{k}_i} \rangle + \mathcal{O} ((\delta k_{\mu})^2) \\
&= 1 - i \delta k_{\mu} \mathcal{A}_\mu(\bm{k}_i) + \mathcal{O} ((\delta k_{\mu})^2) \\
&\approx \left| U_{i,j} \right| e^{-i \delta k_{\mu} \mathcal{A}_\mu(\bm{k}_i)},
\end{aligned}
\end{equation}
with $\mu$ representing the direction from $\bm{k}_i$ to $\bm{k}_j$. 
Thus, the incremental Berry phase along the link is

\begin{equation}
\delta k_{\mu} \mathcal{A}_\mu(\bm{k}_i) \approx -\text{Im}\,\ln\, U_{i,j} = i \ln \frac{U_{i,j}}{\left|U_{i,j}\right|}
\end{equation}

For a closed plaquette boundary $\partial P_i$ with vertices $\bm{k}_1, \bm{k}_2, \dots, \bm{k}_N$ (where $\bm{k}_{N+1} \equiv \bm{k}_1$), the total Berry phase is
\begin{equation}
\begin{aligned}
   \Phi_{i} = \oint_{\partial P_i} \bm{\mathcal{A}} \cdot \mathrm{d}\bm{k}
   &\approx \sum_{l} \mathcal{A}{\mu_l}(\bm{k}_l) \delta k{\mu_l} \\
   &= \sum_{l} \left[ i \ln \frac{U_{l,l+1}}{\left|U_{l,l+1}\right|}\right]  \\
   &= i \ln \left( \prod_{l=1}^{N} \frac{U_{l,l+1}}{\left|U_{l,l+1}\right|} \right)  \\
\end{aligned}
\end{equation}
where the logarithm is taken with the principal branch to ensure $\Phi_{i} \in \left( -\pi, \pi \right]$.
The Chern number is then obtained by summing over all plaquettes:
\begin{equation}
\begin{aligned} 
   C = \frac{1}{2\pi} \sum_{i=1}^{N_p} \Phi_{i}
   = \frac{i}{2\pi} \sum_{i=1}^{N_p} \ln \left( \prod_{l=1}^{N} \frac{U_{l}}{\left|U_{l}\right|} \right) 
\end{aligned}
\end{equation}

This method provides a gauge-invariant and numerically stable way to compute the Chern number. Under a $U(1)$ gauge transformation
$|\psi_{\bm{k}} \rangle \to e^{i\theta(\bm{k})} |\psi_{\bm{k}} \rangle$
, the link variables transform as
$U_{i,j} \to e^{i \left[\theta(\bm{k}_j) -\theta(\bm{k}_i) \right]} U_{i,j}$
and the product over a closed loop is gauge-invariant.


\funding{ This work was funded by National Natural Science Foundation of China (Grant No. 12474144, W.Z., No. 12274354, S.X.), the Zhejiang Provincial Natural Science Foundation of China (Grant No. LR24A040003, S.X.; No. XHD23A2001, S.X.), and Westlake Education Foundation at Westlake University. }

\roles{J.H. and J.D. wrote the initial draft. All authors contributed to discussion. }





\bibliographystyle{iopart-num}
\bibliography{the_bib.bib} 

\end{document}